\documentclass{article}

\usepackage[english]{babel}

\usepackage[letterpaper,top=2cm,bottom=2cm,left=3cm,right=3cm,marginparwidth=1.75cm]{geometry}

\usepackage{mathtools}
\usepackage{amsmath}
\usepackage{graphicx}
\usepackage{adjustbox}
\usepackage[colorlinks=true, allcolors=blue]{hyperref}
\usepackage{natbib}
\usepackage{amsfonts}
\usepackage{subfigure}
\usepackage{hyperref}
\usepackage[ruled,vlined]{algorithm2e}

\title{Two-stage Sampling Design and Sample Selection with the R package R2BEAT}
\date{}

\author{Giulio Barcaroli\footnote{Independent consultant. Email: gbarcaroli@gmail.com},
Andrea Fasulo\footnote{Italian National Institute of Statistics (ISTAT). Email: fasulo@istat.it},
Alessio Guandalini\footnote{Italian National Institute of Statistics (ISTAT). Email: alessio.guandalini@istat.it}, 
Marco D. Terribili\footnote{Italian National Institute of Statistics (ISTAT). Email: terribili@istat.it}}

\begin{document}
\maketitle

\section*{Abstract}
R2BEAT ("R 'to' Bethel Extended Allocation for Two-stage sampling") is an R package for the allocation of a sample.
Besides other software and packages dealing with the allocation problems, its peculiarity lies in facing properly allocation problems for complex sampling designs with multi-domain and multi-purpose aims.
This is common in many official and non-official statistical surveys, therefore R2BEAT could become an essential tool for planning a sample survey.
The package implements the  \cite{tschprow1923optimal} - \cite{neyman1934optimal} method for the optimal allocation of units in stratified sampling, extending it to the multivariate (accordingly to \citeauthor{bethel1989sample}'s proposal (\citeyear{bethel1989sample})), multi-domain and to the complex sampling designs case \citep{falorsi1998principi}. 
The functions implemented in R2BEAT allow the use of different workflows, depending on the available information on one or more interest variables.
The package covers all the phases, from the optimization of the sample to the selection of the Primary and Secondary Stage Units. 
Furthermore, it provides several outputs for evaluating the allocation results.\\

\noindent
\textbf{Keywords} sample survey, multistage, multipurpose, optimal allocation, sample selection. 

\section{Introduction} 
\label{sec:intro}
National Statistical Institutes (NSIs) and other official statistics institutions usually stratify the target population into homogeneous groups, defined by variables. 
Survey data usually benefits from stratification, and sampling error decreases. 
However, from a logistic point of view, the stratified sample could be geographically widespread, entailing such a cost increase in the data collection process.
For solving this issue and to avoid sample dispersing, the two-stage stratified sampling design is often used for planning surveys, mainly the social ones carried out in households. 
This sampling design enables to control of the number of Primary Stage Units (PSUs) selected in the survey. 
For instance, the municipalities or the enumeration areas in which the selected households (Secondary Stage Units, SSUs) belong. 
Controlling the municipalities number remarkable reduces data collection costs, mainly for face-to-face interviews, and avoids logistic problems given by a geographically scattered sample. 
Nevertheless, allocating a two-stage sample among strata can be tricky: usually, households surveys are defined as multipurpose, since they estimate many target variables; moreover, produced estimates are provided for many estimation domains, such as national level, geographical areas, municipality types, etc. 
In this context, the allocation of the whole sample size becomes a multivariate and multi-domain problem. 
It is important to point out that the total size is defined according to three types of constraints: estimates precision, budget, logistic ones, or more likely by a combination of the three. 

Once indicatively defined the whole sample size, intended as the number of SSUs to select and interview, has to be allocated among the strata in which the PSUs population is partitioned. 
Different methods can be used for allocating the sampling units among the strata according to the available information.
The easiest methods are uniform and proportional allocation.
If, however, the values and the variances of some survey target variables are known in each design stratum, from auxiliary sources such as registers or previous survey occasions, then an optimal allocation can be computed.

The idea behind the optimal allocation is that strata with larger sizes and larger variability recorded on the target variables need a larger sample size to provide better estimates. Several publications and packages focus on this aspect. 

{R2BEAT} extends the methodology implemented in Istat's open-source software called \textbf{Mauss-R} \citep{maussr}, which stands for ``Multivariate Allocation of Units in Sampling Surveys", widely used for designing one-stage sample surveys and also in the {SamplingStrata} \citep{barcaroli2014samplingstrata}.

Furthermore, it faces the optimal allocation definition in the two-stage sampling design case. 
Its name stands for {R} "to" Bethel Extended Allocation for Two-stage. The package represents a very specific tool for designing, allocating and selecting the most complex and challenging sample in the context of survey designs. 

Furthermore, {R2BEAT} fills a gap, within the range of statistical software concerning sample size allocation. In fact, several {R} packages are available for allocating a stratified sample, such as {surveyplanning} \citep{surveyplanning}, {PracTools} \citep{practools}, {optimStrat} \citep{optimstrat}, and the already mentioned {Mauss-R}  and {SamplingStrata}, but none of these can compute the optimal allocation among strata in such a complex sampling design context, considering both multivariate and multi-domain case.

In the following paragraph the methodological aspects, underlying the package and its functions, will be presented in detail: the optimal allocation of the sample and its selection will be illustrated. 
In the third paragraph will be shown how to prepare, organize and check the input data needed by the package for allocating the whole sample size among strata and to finally select the units. 
A case study on a synthetic dataset will be used as an example to test the package functions. Finally, the results will be discussed in the concluding remarks.

\section{Methodological aspects}
\label{sec:method}

Sample surveys carried out by National Statistical Institutes and by other institutions have multi-domains and multi-purpose objectives, so they have to provide accurate estimates for different parameters and different domains (i.e. geographical areas such as national, regional, and more).
However, usually, the survey has budgetary constraints, then, they must be carefully planned to provide high-quality estimates for parameters of interest.

A seminal work in this perspective is due to \citet{kish1965survey}. 
While a broad theoretical framework for optimizing surveys by maximizing data quality within budgetary constraints is provided by \citet{biemer2003introduction} and \citet{biemer2010total}.

When designing a multipurpose survey several choices need to be made.
They usually are not trivial, because identifying the best solution for every purpose (i.e. every interest variable for each domain of interest) is challenging.
Usually ``just`` a practical optimum, not the best solution, can be pursued.
The research for the best solution - maximizing data quality within budgetary constraints - may arise conflicts in several 
areas \citep{kish1988multipurpose}.
Among these areas, sample size and the relation of biases to sampling errors are considered the most important because their influence ripples throughout the overall survey. 

This view justifies the care and attention always given in the literature to the optimal sample design \citep{cochran1977sampling,cicchitelli1992campionamento,conti2012campionamento,tille2020sampling}. 
\citet{gonzalez2010optimal} present an interesting overview of the approaches for defining optimal sampling strategies. 

The optimization problem of a sample design is usually dealt with the estimation of a mean (or equivalently of a total) in stratified sampling designs with a fixed sample size. 
The problem of the optimization of stratified sample design can be classified depending on whether stratification is given or also the stratification has to be optimized, before or at the same time of the allocation. 

The {R2BEAT} package solves the optimization problem when the stratification is given and the optimization must be sought in the allocation of sampling units.
Therefore, in the following, we focus just on this situation. 
For more details on the optimization problems when also the stratification has to be optimized see, e.g., \citet{ballin2013joint} and references therein.

\subsection{Optimal allocation}
\label{sec:samplealloc}
Let us consider a population $U$ of size $N$ ($k=1, \dots, N$) partitioned in $H$ subgroups, $U_h$ $(h = 1, \dots, H)$, called strata.
Hence, each stratum contains $N_h$ elements, where $N_h$ is assumed to be known and such as $\sum_{h=1}^{H} N_h = N$. 

The strata can be defined in different ways on the basis of one or more qualitative variables known for all the units in the population.

Then, we assume, at least for the moment, to be interested in investigating the mean of just one $y$ variable in the population $U$,
\begin{eqnarray}
	\label{eq:meanPOP}
	\mu_{y} = \frac{\sum_{k \in \ U} y_k}{N}
\end{eqnarray}
where $y_k$ is the value of the $y$ variable observed on the $k$-th unit in the population $U$.
The $y$ variable could be a quantitative variable or dichotomous, that is $y \in \left\lbrace 0, 1 \right\rbrace$.
Please note that, even when $y$ is a dichotomous variable, expression~\eqref{eq:meanPOP} holds and $\mu_{y}$ is equal to the proportion of units in the population for which $y=1$.

Furthermore, assume we want to estimate $\mu_{y}$ through a probabilistic sample $s$ of size $n$ with the estimator
\begin{eqnarray}
	\label{eq:HT}
	\hat{\bar{Y}} = \frac{\hat{Y}_{HT}}{N} = \frac{\sum_{k \in s} y_k \ d_k}{N}
\end{eqnarray}
where $\hat{Y}_{HT}$ is the Horvitz-Thompson estimator for the total \citep{horvitz1952generalization} in which $d_k$ is the design weight usually equal to the inverse of the first order inclusion probability.

The sample size of a survey, $n$, is usually exogenous information, dictated by budget and, sometimes, by logistic constraints associated to the unit $k$ in the sample.
Then, in practice, the problem comes down to the allocation of the $n$ units in the $H$ strata, such as $\sum_{h=1}^H n_h=n$.

Therefore, let us define 
\begin{eqnarray}
	\label{eq:meanPOPstr}
	\mu_{hy} = \frac{\sum_{U} y_k \ \boldsymbol{1}_h}{N_h}
\end{eqnarray}
the mean of the $y$ in each stratum where $\boldsymbol{1}_h$ is the membership indicator for the unit $k$ in the stratum $h$.

In the same way, expressions~\eqref{eq:HT} can be easily adapted for estimating $\mu_{hy}$, that is
\begin{eqnarray}
	\label{eq:HTh}
	\hat{\bar{Y}}_h = \frac{\hat{Y}_{HT,h}}{N_h} = \frac{\sum_{k \in s_h} y_k \ d_k}{N_h},
\end{eqnarray}
where $s_h$ is the sample in the stratum $h$.
The sampling variance estimator of $\hat{\bar{Y}}_h$ is given by
\begin{eqnarray}
	\label{eq:varHTh}
\widehat{\text{var}} \left(\hat{\bar{Y}}_h \right) =
\frac{1}{n_h - 1} \left(\frac{1}{n_h} - \frac{1}{N_h} \right)
\sum_{k \in h} (y_{hk} - \bar{y}_h)^2,
\end{eqnarray}
where $\bar{y}_h$ is the sample mean of the variable $y$ in the stratum $h$. 

In this perspective, the mean of $y$ in \eqref{eq:meanPOP} can be written also as 
\begin{eqnarray*}
	\mu_{y}= \sum_{h=1}^H \frac{N_h}{N} \mu_{hy} 
\end{eqnarray*}
and, consequently, $\hat{\bar{Y}}$ in \eqref{eq:HT} as
\begin{eqnarray*}
	\hat{\bar{Y}} = \sum_{h=1}^H \frac{N_h}{N} \hat{\bar{Y}}_h.
\end{eqnarray*}
Therefore, the sampling variance estimator for $\hat{\bar{Y}}$ is
\begin{eqnarray*}
	\widehat{\text{var}} \left( \hat{\bar{Y}} \right) = \sum_{h=1}^H \left(\frac{N_h}{N} \right)^2 \widehat{\text{var}} \left(\hat{\bar{Y}}_h \right).
\end{eqnarray*}

When there is no information on $y$, the sample size to be allocated to each stratum, $n_h$, can be assigned by performing uniform or proportional allocation.

Uniform allocation assigns an equal number of sampling units to each stratum, that is 
\begin{eqnarray*}
	n_h^{UNIF}=\frac{n}{L}.
\end{eqnarray*}

More often, we want the sample size assigned to strata in the sample to be proportional to the sizes of the strata in the population, that is
\begin{eqnarray*}
	n_h^{PROP}=n \ \frac{N_h}{N}
\end{eqnarray*}
where $N_h/N$ is the weight of the stratum in the population with $\sum_{h=1}^L N_h/N=1$.
If the size is the same for all strata ($N_1=\dots=N_h=\dots=N_L=N/L$), $n_h^{PROP}$ comes down to $n_h^{UNIF}$. 

When there is information in the population strata on $y$ and in particular on its variance, $S_{yh}^2$, a more favourable allocation can be performed.
Alternatively, it is possible to consider also a proxy variable highly correlated with $y$.
In this case, \cite{tschprow1923optimal} demonstrated that the optimal allocation can be obtained by giving 
\begin{eqnarray*}
	n_h^{OPT}=n \frac{\frac{N_h}{N} \ \sqrt{S_{yh}^2}}{\sum_{h=1}^L \frac{N_h}{N} \ \sqrt{S_{yh}^2}}  \end{eqnarray*}
However, this result is better known as \citeauthor{neyman1934optimal} allocation by the namesake author that in \citeyear{neyman1934optimal} published the same result.
The rationale behind the optimal allocation is that strata with more weight and in which $y$ has much more variability need much more observations for reaching better estimates. 
If the variance is the same in all the strata ($S_1^2=\dots=S_h^2=\dots=S_L^2$), $n_h^{OPT}$ comes down to $n_h^{PROP}$. 

As evidence, the computation of the population variance is a crucial point in the optimal allocation.
A distinction between the types of variables and the sources from which they can be obtained is needed.
When $y$ is a dichotomous variable available from a population register, its population variance can be computed as
\begin{eqnarray}
	\label{S2dicho}
	S_{yh}^2 = p_h \times (1 - p_h)
\end{eqnarray}
where $p_h$ is the proportion of units with $y=1$ in the population strata. 
In the case of a quantitative variable, $S_{yh}^2$ is equal to 
\begin{eqnarray*}
	\label{S2quant_register}
	S_{yh}^2  = \frac{\sum_{k \in U_h} \left( y_k - \mu_{yh} \right)^2}{N_h}.
\end{eqnarray*}

When there is no population register, information on the variability can be obtained from a sample survey or a pilot survey previously carried out. 
Let us assume to have collected the $y$ variable, or at least its proxy variable, on a sample $s^*$. 
Then, (\ref{S2dicho}) can be computed just by replacing $p_h$ with 
\begin{eqnarray}
	\label{eq:estp}
	\hat{p}_h = \frac{\sum_{k \in s_h^*} y_k \ w_k}{\sum_{k \in s_h^*} w_k},     
\end{eqnarray}
that is the related estimate for each stratum obtained from the sample $s^*$.
In \eqref{eq:estp} $w_k$ is the sampling weight associated with the unit $k$ in the sample $s^*$. 

Instead, when $y$ is a quantitative variable,
\begin{eqnarray*}
	\label{S2quant_survey}
	S_{yh}^2  =  \hat{M}_h^2 - \hat{\bar{Y}}_h^{2}
\end{eqnarray*}
where 
\begin{eqnarray*}
	\hat{M}_h^2 = \frac{\sum_{k \in s_h^*} y_{k}^2 \ w_k}{N} \hspace{0.5cm} \text{and} \hspace{0.5cm}
	\hat{\bar{Y}}_i = \frac{ \sum_{k \in s_h^*} y_{k} \ w_k}{N}
\end{eqnarray*}
are the quadratic mean and the arithmetic mean estimated on the sample $s^*$ in the $h$-th stratum, respectively. 

Sometimes, collecting data on units belonging to different strata can have different costs for the difficulties in reaching them (e.g. strata are altitude zone) or the need of using different data collection modes.
Therefore, it is advisable to define the allocation by taking into account also the unit cost and the budget constraints.

The global cost of the survey can be defined as
\begin{eqnarray*}
	C=c_0 + \sum_{h=1}^L n_h \ c_h,
\end{eqnarray*}
where $c_0$ is the fixed cost (not dependent on the sample size) and $c_h$ is the unit cost for collecting data on one unit belonging to stratum $h$.
Then, the optimal allocation under budget constraints is given by
\begin{eqnarray*}
	n_h^{\bar{OPT}}=(C-c_0) \frac{\frac{\frac{N_h}{N} \ \sqrt{S_{yh}^2}}{\sqrt{c_h}}}{\sum_{h=1}^L \frac{N_h}{N} \ \sqrt{S_{yh}^2} \sqrt{c_h}}.
\end{eqnarray*}
If $c_1=\dots=c_h=\dots=c_L=1$ and $c_0=0$, the global cost amounts to the sample size ($C=n$).

The optimal allocation for just one $y$ variable is of little practical use unless the various variables under study are highly correlated. 
This is because an allocation that is optimal for one characteristic is generally far from being optimal for others.

Therefore, several works have been devoted to solving the problem when more than one variable of interest has to be measured on each sampled unit.
All the contributions can be classified into two main approaches: the ``average variance" and convex programming.

The methods under the ``average variance" approach consist of defining a weight for each variable to consider, computing a weighted average of the stratum variance and finding the optimal allocation on the ``average variance" which results.
They are computationally simple, intuitive and can be solved under fixed cost assumption.
However, the choice of the weights is completely arbitrary and the optimal properties are not clear \citep[see, e.g.,][for more details]{dalenius1953multi, yates1960sampling,folks1965optimum, hartley1965multiple,kish1976optima}.

Instead, the other approach includes methods that use convex programming to find the minimum cost allocation when the variances of all the sampling variables to consider satisfy fixed constraints.     
The obtained allocation is actually optimal, but sometimes it can exceed the budgetary constraints \citep[see, e.g.,][for more details]{dalenius1957sampling,yates1960sampling,kokan1963optimum,hartley1965multiple,kokan1967optimum,chatterjee1968multivariate,chatterjee1972study,huddleston1970optimal,bethel1985optimum,chromy1987design,falorsi1998principi,stokes2004using,choudhry2012sample,kozak2007modern,kozak2008stratified}. 

The most important method in the convex programming approach is the Bethel algorithm \citep{bethel1989sample} which extends the Neyman allocation to the multivariate case.

In particular, when we are interested in investigating the mean of more than one $y$ variable (quantitative or dichotomous), namely $y_1, \dots, y_i, \dots, y_J$, the optimal allocation problem reduces, in practice, to a minimum optimization problem of a convex function under a set of linear constraints
\begin{eqnarray}
	\label{eq:optprob}
	\left\lbrace
	\begin{array}{l}
		C = \text{min} \\
		\widehat{CV} \left(\hat{\bar{Y}}_{i,h}\right) \leq 
		\delta \left(\hat{\bar{Y}}_{i,h} \right)
		\hspace{0.5cm} \begin{array}{c}
			i = 1, \dots, J \\
			h = 1, \dots, L 
		\end{array}
	\end{array}
	\right.
\end{eqnarray}
where $C$ is the global cost of the survey and $\hat{CV} \left(\hat{\bar{Y}}_{i,h}\right)$ is the estimate of the relative error. 
The estimate of the relative error,
\begin{eqnarray}
	\label{eq:CVHTh}
	\widehat{CV} \left(\hat{\bar{Y}}_{i,h} \right) = \frac{\sqrt{\widehat{\text{var}} \left(\hat{\bar{Y}}_{i,h} \right)}}{\hat{\bar{Y}}_{i,h}}, 
\end{eqnarray}
is the ratio between the estimate of the sampling variance for the mean estimator of $y_i$ variable ($i=1, \dots, J$) in  the stratum $h$ given by expression~\eqref{eq:varHTh} and the related estimate.
In this case, $\widehat{CV} \left(\hat{\bar{Y}}_{i,h}\right)$ is called
expected errors and it must be less than or equal to the precision constraints defined by the user or by regulation, $\delta \left(\hat{\bar{Y}}_{i,h} \right)$.

\cite{bethel1989sample} demonstrates that the solution to this optimization problem exists and can be obtained through an algorithm that applies the Lagrangian multipliers method.
The solution is a continuous solution, then it must be rounded to provide an integer stratum sample size.
The rounding  clearly causes some deviations from the solution that, however, do not affect its optimality \citep{cochran1977sampling}.
The Bethel algorithm is very similar to the Chromy algorithm \citep{chromy1987design}.
However, it is preferable because, even if the Chromy algorithm is simpler, there is no proof that it converges if a solution exists.

The same framework works to deal also with the multi-domain problem.

Usually, estimates of a survey are disseminated for
the whole population and sub-domains, for instance for geographical areas, but not only.
Then, it is useful to define the optimal allocation also taking into account these outcomes of the survey.

Sub-domain estimation is actually a long-established theory \citep{SSW:03}. 
Expressions~\eqref{eq:meanPOP} can be easily adapted just by introducing the sub-domain membership indicator variable, $\boldsymbol{1}_{k,d}$, which is equal to 1 for all the unit $k$ in the domain $d$ and 0 otherwise, that is
\begin{eqnarray*}
	\label{eq:meanPOP_dom}
	\mu_y^d = \frac{\sum_{U_d} y_k \ \boldsymbol{1}_{k,d}}{N_d}
\end{eqnarray*}
where $N_d$ is the population size in the domain $d$ ($d=1, \dots, D$).
It is important to point out, that domains must be an aggregation of strata and they do not have to cut the strata.
Then, it is sufficient to consider the domain estimates in the minimum optimization problem in \eqref{eq:optprob} and use the Bethel's algorithm for deriving the multivariate allocation in the multi-domain case. 

However, in official statistics, especially for household surveys, two-stage sampling designs are usually adopted.

Two-stage sampling is based on a double sampling procedure: one on the primary stage units (PSUs) and another on the second stage units (SSUs).
For instance, in the household survey, the PSUs are the municipalities that are firstly selected.
Then, in each selected municipality, a sample of households - the $SSU$ - can be selected.

Two-stage sampling permits more complex sampling strategies and, moreover, it helps in the organization and cost reduction of data collection, because it reduces the interviewer's travels.
However, this economic saving is paid off with a loss of efficiency of the estimates.
In fact, each additional stage of selection usually entails an increase of the sampling variance of the mean estimator.
This increase can be assessed by the design effect ($deff$) that measures how much the sampling variance of $\hat{\bar{Y}}_{i}$, under the adopted sampling design ($des$), is inflated with respect to a simple random sample ($srs$), with the same sample size.
An estimate of the design effect, under  can be given by the expression:
\begin{eqnarray*}
	\label{eq:deff}
	deff \left( \hat{\bar{Y}}_{i} \right) & = & \frac{\widehat{\text{var}} \left( \hat{\bar{Y}}_{i}\right)_{des}}{\widehat{\text{var}} \left(\hat{\bar{Y}}_{i}\right)_{srs}}
\end{eqnarray*} 
While a rough approximation of the $deff$ can be obtained when the clusters have the same sample size and the same inclusion probability \citep{cicchitelli1992campionamento},
\begin{eqnarray}
	\label{eq:deff1}
deff \left( \hat{\bar{Y}}_{i} \right) & = & 1+\rho_i \ (b-1)
\end{eqnarray} 
where $b$ is the average cluster (i.e. PSU) size in terms of the final sampling units and $\rho_i$ is the intra-class correlation within the cluster (PSU) for the variable $y_i$ $(i=1, \dots, J)$. 

The intra-class correlation provides a measure of data clustering in PSUs and SSUs. 
In general, if $\rho_i$ is close to 1, the clustering is high and it is convenient to collect only a few units in the cluster. 
On the contrary, if $\rho_i$ is close to 0, the collection of units from the same cluster does not affect the efficiency of the estimates. 

Also for computing $\rho_i$, we can distinguish whether a population register in which the $y_i$ variables ($i=1,\dots,J$), or at least their proxies, are available or not.
In the former case, a good approximation is given by the expression 
\begin{eqnarray}
	\label{eq:rho_pop}
    \rho_i=1 - \frac{D_{w_i}}{D_{y_i}}  
\end{eqnarray}
where 
\begin{eqnarray*}
	D_{w_i} & = & \sum_{\ell=1}^L \sum_{k=1}^{N_\ell} \left( y_{i,k} - \mu_{y_{i,\ell}} \right)^2
\end{eqnarray*}
and
\begin{eqnarray*}
	D_{y_i} & = &\sum_{k \in U} \left(y_{i,k} - \mu_{y_i}\right)^2
\end{eqnarray*}
are the deviance within clusters and the global deviance of the $y_i$ variable, respectively.
Remember that $D_{y_i}= D_{w_i} + D_{b_i}$, where
\begin{eqnarray*}
	D_{b_i} = \sum_{\ell=1}^L N_\ell \left(\mu_{yi,\ell} - \mu \right)^2,
\end{eqnarray*}
is the deviance between clusters.
Therefore, $0 \leq \rho_i \leq 1$. 

Instead, $\rho_i$ can be estimated from a sample with the expression~\eqref{eq:deff1}
\begin{eqnarray}
	\label{eq:rho_est}
	\hat{\rho}_{i} = \frac{deff_i - 1}{b-1}.
\end{eqnarray}

Here we consider, directly, a more general expression for the estimate of the $deff$ in terms of the intra-class correlation coefficient.
This expression refers to a typical situation in household surveys where PSUs are assigned to Self-Representing (SR) strata, that is they are included for sure in the sample, or to Not-Self-Representing (NSR) strata, where they are selected by chance.
In practice, this assignment is usually performed by comparing the measure of the size of PSUs to the threshold:
\begin{equation}
	\label{eq:threshold}
	\lambda = \frac{\bar{m} \ \Delta}{f}
\end{equation}
where $\bar{m}$ is the minimum number of SSUs to be interviewed in each selected PSU, $f=n/N$ is the sampling fraction and $\Delta$ is the average dimension of the SSU in terms of elementary survey units.
Then, $\Delta$ must be set equal to 1 if, for the survey, the selection units are the same as the elementary units (that is, household-household or individuals-individuals), whereas it must be set equal to the average dimension of the households if the elementary units are individuals, while the selection units are the households. 

PSUs with a measure of size exceeding the threshold are identified as SR, while the remaining PSUs are identified as NSR.

Then, the extended expression of $deff$ \citep[see among the others][]{rojas2016estrategias} is
\begin{equation} 
	\label{eq:deff_ext}
	deff \left(\hat{\bar{Y}}_{i}\right) 
	=  \frac{N_{SR}^2}{n_{SR}} \left[ 1 + \left( \rho_{i,SR} \  (b_{SR} - 1 \right) \right] +
	\frac{N_{NSR}^2}{n_{NSR}} \left[ 1 + \left(\rho_{i,NSR} \ (b_{NSR} - 1 \right) \right]
\end{equation}
where, for $SR$ and $NSR$ strata, 
\begin{itemize}
	\item $N_{SR}$ and $N_{NSR}$ are the population sizes;
	\item $n_{SR}$ and $n_{NSR}$ are the sample sizes;
	\item $\rho_{i,SR}$ and $\rho_{i,NSR}$ the intra-class correlation coefficients for the variable $i$ ($i=1, \dots, J)$;
	\item $b_{SR}$ and $b_{NSR}$ are the average PSU size in terms of the final sampling units.
\end{itemize} 
Of course, if there are no SR strata the expression \eqref{eq:deff1} recurs. 
The design effect is equal to 1 under the $srs$ design and increases for each additional stage of selection, due to the intra-class correlation coefficient which is, usually, positive.

The intra-class correlation coefficient for NSR can be derived with expression ~\eqref{eq:rho_pop} or \eqref{eq:rho_est} whether population register data are available or not.
While it is not necessary to compute the intra-class correlation coefficient for SR strata because just one PSU is selected and the intra-class correlation is 1 by definition.

Therefore, under a two-stage sample design for determining the optimal allocation, the number of PSUs and SSUs must be determined.

The solution has been proposed by \cite{falorsi1998principi} in a paper published in Italian and it is obtained with an iterative use of the Bethel algorithm.
In fact, at the first iteration, the Bethel algorithm is applied. 
The optimal allocation for a stratified simple sampling design is obtained. 
Then, this allocation is used to update the threshold in \eqref{eq:threshold} and the design effect in \eqref{eq:deff_ext}.
A new design effect is computed and used in turn to inflate the $S_h^2$ (or equivalently $\hat{S}_h^2$).
It is used as input in the next iteration in which the Bethel algorithm is used again.
The obtained allocation is used again to update the threshold and the design effect, and a new allocation is found.
The process is iterated until when the difference between two consecutive iterations is lower than a predefined threshold. \\ \\ \\

\begin{algorithm}[h]
	\SetAlgoLined
	\SetKwData{Left}{left}
	\SetKwData{This}{this}\SetKwData{Up}{up}
	\SetKwFunction{Union}{Union}
	\SetKwFunction{FindCompress}{FindCompress}
	
	\SetKwInOut{Input}{Input}
	\SetKwInOut{Output}{Output}
	\BlankLine
	\Input{
		\BlankLine
		a. precision constraints in terms of CV\;
		b. information on sampling strata (mean and stdev of target variables, N, ...)\;
		c. information on previous design: deff, effst, rho \;
		d. information on PSUs in sampling strata (measure of size)\;
		e. minimum number of SSUs per PSU\;
	}
	\BlankLine
	\Output{
		\BlankLine
		a. for each stratum: number of PSUs and SSUs to be selected\;
		b. expected CVs for target estimates\;
		c. item sensitivity of the solution\;
	}
	\BlankLine
	\nlset{REM} First iteration\;\label{first}
	1. input deff is used to inflate standard deviations of target variables in sampling strata\;
	2. optimal allocation of SSUs in sampling strata is obtained by applying the Bethel algorithm as if it were a one-stage sampling design\;
	3. the number of PSUs is determined on the basis of the minimum number of SSUs per PSU\;
	4. the threshold for determination of self-representing PSUs is calculated\;
	5. new deff is calculated and used to update the standard deviations of target variables in sampling strata\;
	
	\nlset{REM} Next iterations\;\label{iterations}
	\While{not convergence} {
		1. optimal allocation of SSUs in sampling strata is obtained by applying the Bethel algorithm\;
		2. the number of PSUs is determined on the basis of the minimum number of SSUs per PSU\;
		3. the threshold for determination of self-representing PSUs is calculated\;
		4. new deff is calculated an used to update standard deviations of target variables in sampling strata\;
		5. the iteration stops if\\
		\begin{itemize}
			\item[a.] the difference between the sample sizes of two iterations is lower than 5 (default value) \textit{or}
			\item[b.] the maximum of defts (square root of deffs) largest differences is lower than 0.06 (default value) \textit{or}
			\item[c.] the number of iterations is higher than 20 (default value)\;
		\end{itemize}  
	}
	
	\caption{R2BEAT optimal allocation of PSUs and SSUs in sampling strata}
	\label{algorithm}
\end{algorithm}

However, as pointed out by \cite{waters1987optimal}, different combinations yield the same variance and can satisfy the precision constraints, $\delta \left( \hat{\bar{Y}}_{i,h} \right)$.
The optimal solution strongly depends on the budgetary constraints that limit the $SSU$s and the data collection organization that influences the maximum number of $PSU$s that can be managed.

All this discussion holds when you want to use the $HT$ estimator.   
But, currently, the most applied estimator for the NSIs survey is the calibrated estimator \citep{deville1992calibration,sarndal2007calibration,devaud2019deville}.
The calibrated estimator, through the use of auxiliary variables, usually provides better estimates than $HT$.
Then, it can be useful to take into account, since the allocation phase, also be the impact on the estimates of an estimator different from the $HT$ estimator.
This can be done by inflating the $S_{yh}^2$ with the estimator effect and following the procedure explained above.
An estimate of the estimaror effect ($effst$) is given by 
\begin{equation}
	\label{effst}
	effst (\hat{\bar{Y}}_{i}) = \frac{\text{var} \left(\hat{\bar{Y}}_{i} \right)}{\text{var} \left(\hat{\bar{Y}}_{i,_{HT}} \right)}.
\end{equation}
It measures how much the sampling variance of the applied estimator under the adopted design is inflated or deflated with respect to the sampling variance of the $HT$ estimator, on the same sample design.

\subsection{Sample selection}
\label{sec:samplesel}
Once the optimal allocation is defined, the selection of sampling units must be performed.

In the case of a stratified two-stage sampling design two sampling selections need to be done: one for PSUs and one for SSUs.

In each stratum, the PSUs are split into SR and NSR according to a size threshold \eqref{eq:threshold}. 
PSUs with a measure of size exceeding the threshold are identified as SR, included for sure in the sample and each of them constitutes an independent sub-stratum. 
Therefore, the probability that they are included in the sample (inclusion probability,  $\pi_I$) is always equal to 1.
However, it can happen that no one PSU has a measure of size higher than the threshold.

The remaining PSUs, NSR-PSUs, are ordered by their measure of the size and divided into finer strata (\textit{sub-strata}) whose sizes are approximately equal to the threshold multiplied by the number of PSUs to be selected in each stratum.
In this way, sub-strata are composed of PSUs having size as homogeneous as possible. 

The PSUs in each stratum can be selected in different ways.
However, the selection of a fixed number of PSUs per stratum is usually carried out with Sampford's method (unequal probabilities, without replacement, fixed sample size).
Then, the inclusion probability of the generic $\ell$-th NSR-PSU, is 
\begin{equation*}
	\pi_I=\frac{N_h}{m \ M_{h\ell}}
\end{equation*}
where $N_h$ is the measure of size in the sub-stratum $h$-th, $m$ is the number of NSR-PSUs to be selected in the sub-stratum and $M_{h\ell}$ is the measure of size in the $\ell$-th PSU in the sub-stratum $h$. 

Finally, the SSUs must be drawn in the selected PSU.
Also in this case the SSU can be selected in different ways.
In most cases, they are selected through a systematic sampling design that shares several properties with the $srs$.
Then, the inclusion probability for the second stage is equal to
\begin{equation*}
	\pi_{II}=\frac{n_{h\ell}}{M_{h\ell}}
\end{equation*}
where $n_{h\ell}$ is the number of SSUs to be selected in the $\ell$-th PSU in the $h$-th sub-stratum.

Then, the design weight for the unit $k$ in the $h$-th strata in the $\ell$-th PSU is equal to the inverse of the product of the first stage and the second stage inclusion probabilities, 
\begin{equation*}
	d_k = \frac{1}{\pi_{I}} \frac{1}{\pi_{II}}.     
\end{equation*}
The design weights sum up to the population size, $\sum_{k \in s} d_k = N$, and are almost constant in each stratum, which means that the sample is self-weighting. \\ \\ \\

	\begin{algorithm}[H]
	\SetAlgoLined
	\SetKwData{Left}{left}
	\SetKwData{This}{this}\SetKwData{Up}{up}
	\SetKwFunction{Union}{Union}
	\SetKwFunction{FindCompress}{FindCompress}
	
	\SetKwInOut{Input}{Input}
	\SetKwInOut{Output}{Output}
	\BlankLine
	\Input{
		\BlankLine
		a. The output of the allocation step (function \texttt{beat.2st}) (universe of PSUs, measure of PSUs, number of PSUs and SSUs to be selected in each stratum, threshold)\;
	}
	\BlankLine
	\Output{
		\BlankLine
		a. universe of PSUs with stratum, sub-stratum, PSU first order inclusion probability, PSU weight, flag sample, and number of SSUs to be selected in each PSU\;
		b. sample of PSUs (flag sample=1) with stratum, sub-stratum, PSU first order inclusion probability, PSU weight, number of SSUs to be selected in each PSU\;
		c. statistics related to the sample of PSUs at stratum level\;
	}
	\BlankLine
	\nlset{REM} creation of \textit{sub-strata} and selection of PSUs\;
	1. in each stratum, PSUs are sorted in descending order according to their measure of size\;
	2. the measure of size of PSUs are compared with the threshold\;
	3. PSUs with a measure of size exceeding the threshold are identified as SR, included for sure in the sample and constitutes an independent sub-stratum\;
	4. the remaining PSUs, NSR-PSUs, are ordered in decreasing way by their measure of the size and aggregated into finer strata (\textit{sub-strata})\;
	5. \textit{sub-strata} are created adding PSUs (still in descending order of measure of size) for which the sum of the measure of size of the \textit{sub-strata} is approximately equal to the threshold multiplied by the number of PSUs to be selected in each stratum\;
	6. in each \textit{sub-stratum} a fixed number of PSUs per stratum are usually selected with Sampford's method (unequal probabilities, without replacement, fixed sample size)\;
	
	\caption{R2BEAT selection of PSUs}
	\label{algorithm2}
\end{algorithm}

\section{Structure of the package} 
The {R2BEAT} package provides functions for drawing complex sample designs using an optimal allocation also performing  the selection of the PSUs and SSUs.
To install the latest release version of {R2BEAT} from CRAN, type \textbf{install.packages("R2BEAT")}
within {R}. 
The current development version can be downloaded and installed from GitHub by executing

\textbf{devtools::install\_github("barcaroli/R2BEAT")}.

This section provides an introduction to the structure and functions associated with the package while the next section will present examples of its specific use.

The workflow to draw and select a complex sample using {R2BEAT} is: (1) prepare the input data, (2) check the input data,
(3) define the design and obtain the allocation, and (4) select the final sample units.

\subsection{Prepare the input data}
\label{sec:prepinput}
As it will be illustrated in detail in the next sub-sections the {R2BEAT} package provides functions to define one-stage stratified sample design (\textbf{beat.1st}) and two-stage stratified sample design (\textbf{beat.2st}). The preparation of the input dataset changes whether the former or the latter sample design will be adopted.

In the case of a multivariate optimal allocation for different domains in a stratified one-stage sample design, the function \textbf{beat.1st} can be used. 
The inputs required by this function are two, 	
a data frame containing survey strata information (\textbf{stratif}) and a data frame of expected CV for each domain and each variable (\textbf{errors}). No functions to prepare these inputs are provided by the package but is possible to follow the example dataset \textbf{stratif} and \textbf{error} to properly create the input datasets for the function \textbf{beat.1st}.

In the case of a two-stage design, two functions are provided by the package to help in the creation of the input data for the function \textbf{beat.2st}. 
The functions are two because two different scenarios are possible, depending on the initial information available:

1. Only the sampling frame is available, no previous rounds of the survey have been carried out. In this scenario, a strict condition on the information content of the sampling frame must hold: values of the sample target surveys (or of their proxy correlated variables) are available for each unit in the frame. This can be accomplished by considering the previous census, or by using administrative registers. In this scenario, the function \textbf{prepateInputToAllocation1} can be used to create the input dataframes \textbf{stratif}, \textbf{rho}, \textbf{deft}, \textbf{effst}, \textbf{des\_file} and \textbf{psu\_file}.  

2. Together with a sampling frame containing the units of the population of reference, also a previous round of the sampling survey to be planned is available. The \textbf{prepateInputToAllocation2} produces the same outputs of \textbf{prepateInputToAllocation1}, but it requires the design and/or calibrated objects of the previous sample survey, obtained using the {ReGenesees} package \citep{zardetto2015regenesees}. 

The function \textbf{sensitivity\_min\_SSU} allows analyzing the different results in terms of first stage size (number of PSUs) and second stage size (number of SSUs), obtained when varying the values of the minimum number of SSUs to be selected in each PSU.

To check the coherence between the estimated population in the strata (\textbf{stratif}) and the population calculated by the PSUs dataset (\textbf{des\_file}), the function \textbf{check\_input} is provided to the users. 
This function compares the strata sizes giving information about the differences and replacing the estimated stratum size with the stratum population calculated by the PSUs dataset.

\subsection{Defining the design and determining the allocation} \label{sec:alloc}

As already introduced, the package allows performing the optimal allocation for both one-stage and two-stage stratified sampling

The first one is implemented within the function \textbf{beat.1st} and computes a multivariate optimal allocation for different domains in one-stage stratified sample design.
As described in section~\ref{sec:prepinput}, in a one-stage stratified sample design there are only two inputs to be provided to \textbf{beat.1st}: the dataframes \textbf{stratif} and \textbf{errors}. Besides these two mandatory inputs, it is also possible to indicate the minimum number of sampling units to be selected in each stratum, by default set equal to 2.

The function \textbf{beat.2st} performs the same multivariate optimal allocation for different domains considering stratified two-stage design. 
Together with the input data \textbf{stratif} and \textbf{errors} other mandatory input are: 

\begin{itemize}
	\item \textbf{des\_file}: dataframe containing a row per each stratum, with information on total population, the values of the \textbf{delta} parameter (equal to the mean number of final SSUs contained in clusters to be selected, for instance, the mean number of individuals in a household), and the minimum number of SSUs to be selected in each PSU;
	\item \textbf{psu\_file}: dataframe containing information on each PSUs (identifier, stratum, measure of size).
	\item \textbf{rho}: dataframe contains a row per each stratum with the intra-class correlation coefficient both for self representing and non-self representing PSUs.
\end{itemize}

Is also possible to provide optional information about:

\begin{itemize}
	\item \textbf{deft\_start}: dataframe containing a row per each stratum with the starting values for the square root of the design effect in the stratum of each variable of interest.
	\item \textbf{effst}: dataframe containing a row per each stratum with the estimator effect for each variable of interest.
\end{itemize}

The functions \textbf{beat.1st} and \textbf{beat.2st} produce lists with respectively 4 and 9 items. 

The \textbf{beat.1st} output contains:
\begin{enumerate}
	\item \textbf{n}: a vector with the optimal sample size for each stratum;
	\item \textbf{file\_strata}: a dataframe corresponding to the input dataframe \textbf{stratif} with the $n$ optimal sample size column added;
	\item \textbf{alloc}: a dataframe with optimal (\textbf{ALLOC}), proportional (\textbf{PROP}), equal (\textbf{EQUAL}) sample size allocation;
	\item \textbf{sensitivity}: a data frame with a summary of planned coefficients of variation (\textbf{Planned CV}), the expected ones under the given optimal allocation (\textbf{Actual CV}), and the sensitivity at 10\% for each domain and each variable. Sensitivity can be a useful tool to help in finding the best allocation, as it provides a hint of the expected sample size variation for a 10\% change in planned CVs.
\end{enumerate}

Together with the previous outputs, the function \textbf{beat.2st} produces also:
\begin{enumerate}
	\item \textbf{iterations}: a dataframe that for each iteration of the Bethel algorithm provides a summary with the number of PSUs (\textbf{PSU\_Total}), distinguished between SR (\textbf{PSU\_SR}) and NSR (\textbf{PSU\_NSR}), plus the number of SSUs;
	\item \textbf{planned}: a dataframe with the planned coefficients of variation for each variable in each domain.
	\item \textbf{expected}: a dataframe with a summary of expected coefficients of variation under the given optimal allocation for each target variable in each domain;
	\item \textbf{deft\_c}: a dataframe with the design effect for each variable in each domain in each iteration. Note that \textbf{DEFT1\_0 - DEFTn\_0} is always equal to 1 if \textbf{deft\_start} is NULL; otherwise it is equal to \textbf{deft\_start}. While \textbf{DEFT1 - DEFTn} are the final design effect related to the given allocation.
	\item \textbf{param\_alloc}: a vector with a resume of all the parameters given for the allocation.
\end{enumerate}

\subsection{Sample units selection} \label{sec:selection}
Once the allocation for the primary and secondary sampling stage units has been defined, it is possible to use two functions for the selection of the final sampling units.

The function \textbf{select\_PSU} allows the users to select the PSUs allocated in each stratum, using the Sampford method, as implemented by the \textbf{UPsampford} function of R package \textbf{sampling} \citep{Rsampling}. 

The input of this function is the output of the \textbf{beat.2st} function.

The output of the function is a list containing the following items:

\begin{enumerate}
	\item \textbf{universe\_PSU}: a dataframe that reports the whole universe of PSUs, with the inner strata formed for the selection;
	\item \textbf{sample\_PSU}: a dataframe containing the selected PSUs, with the indication, for each of them, of how many SSUs must be selected;
	\item \textbf{PSU\_stats}: a table containing summary information on selected PSUs.
\end{enumerate}

In the last step, the selection of a sample of SSUs has to be carried out. The function \textbf{select\_SSU} allows selecting a sample of SSUs from the population frame, based on the SSUs allocated to each selected PSUs. 

The input datasets are two:
\begin{enumerate}
	\item \textbf{df}: the dataframe containing the final sampling units;
	\item \textbf{PSU\_sampled}: the dataframe containing selected PSUs, corresponding to the second item of the output of the \textbf{select\_PSU}function.
\end{enumerate}

The function \textbf{select\_SSU} returns a dataframe containing the  selection of the \textbf{df} dataframe, enriched with information about the first stage inclusion probability, the second stage inclusion probability, the final inclusion probability (the product of the first stage and the second stage inclusion probabilities) and the design weights.

\section{Illustrative examples} \label{sec:example}

To illustrate how to implement workflows making use of \textbf{R2BEAT} functions, we will consider two scenarios, depending on the initial setting:

\begin{enumerate}
	\item only the sampling frame is available, no previous rounds of the survey have been carried out;
	\item together with a sampling frame containing the units of the population of reference, also a previous round of the sampling survey to be planned is available;
\end{enumerate}

In both cases, we assume that the sampling frame contains information on the final sampling units, together with the indication of the PSUs to which each unit belongs.
In the first scenario, a stricter condition on the information content of the sampling frame must hold: values of the sample target surveys (or of their proxy correlated variables) must be available for each unit in the frame. This can be accomplished by considering a previous census, or by imputing values using predictive models. 
In the following paragraphs, we will show only a subset of the code necessary to produce the final results, the relevant part of it\footnote{In order to reproduce the processing related to these examples, datasets and R scripts are downloadable from the link  \href{https://github.com/barcaroli/R2BEAT\_workflows}{https://github.com/barcaroli/R2BEAT\_workflows}.}.

\subsection{Scenario 1 workflow}

In this scenario, it is assumed that a sampling frame is available. We consider a frame (\textbf{pop.RData}), containing 2,258,507 units: 

\begin{verbatim}
	  region province municipality id_hh id_ind stratum stratum_label sex   cl_age
	1  north  north_1            1    H1      1   12000     north_1_6   1  (24,34]
	2  north  north_1            1    H1      2   12000     north_1_6   2  (64,74]
	3  north  north_1            1    H1      3   12000     north_1_6   1 (74,112]
	4  north  north_1            1   H10      4   12000     north_1_6   2  (44,54]
	5  north  north_1            1  H100      5   12000     north_1_6   1  (34,44]
	6  north  north_1            1  H100      6   12000     north_1_6   1  (54,64]
	...
	  active income_hh unemployed inactive
	1      1  30487.75          0        0
	2      1  30487.75          0        0
	3      0  30487.75          0        1
	4      1  21755.68          0        0
	5      1  29870.56          0        0
	6      1  29870.56          0        0
	...
\end{verbatim}	

covering a (synthetic) population of reference, with basic information (geographical and demographic variables: 

\begin{itemize}
	\item region: the NUTS2 identifier;
	\item province: the NUTS3 identifier;
	\item municipality: identifier of the municipality, that plays the role of the PSU identifier;
	\item id\_hh: the household identifier;
	\item id\_ind: the individual identifier;
	\item stratum and stratum\_label: identifier of the initial strata (provinces);
	\item sex and cl\_age: demographic information on individuals.
\end{itemize}

together with information that is related to the sampling survey we want to design:

\begin{itemize}
	\item \textbf{active}, \textbf{inactive}, \textbf{unemployed}: binary variables indicating the occupation status of the individual; 
	\item \textbf{income\_hh}: household income.
\end{itemize}

We suppose that the values of these variables have been made available by a different source (for instance a census) or by predicting them with a model-based approach. In any case the uncertainty related to these values should be taken into account, by correctly evaluating the anticipated variance related to the models used for the predictions when producing the \textbf{strata} dataset \citep[][p. 59]{Baillargeon+Rivest:2012}.

Anyway, in the following, we will not consider this issue, as we want only to illustrate how it is possible to automatically derive all the inputs required by the next steps.

\subsubsection{Step 1: preparation of the inputs for the optimal sample design}

The function \textbf{prepareInputToAllocation1} allows preparing all the inputs required by the optimal allocation step under this first scenario. This function requires the attribution of values to the following parameters:

\begin{itemize}
	\item \textbf{samp\_frame} 
	\item \textbf{id\_PSU} 
	\item \textbf{id\_SSU} 
	\item \textbf{strata\_var} 
	\item \textbf{target\_vars} 
	\item \textbf{deff\_var} 
	\item \textbf{domain\_var}
	\item \textbf{delta} (average dimension of the SSU in terms of elementary survey units)
	\item \textbf{minimum} (minimum number of SSUs to be interviewed in each selected PSU)
\end{itemize}

About the values of these parameters, the choices are almost always driven by the content and structure of the sampling frame, except for \textbf{minimum}. In order to orientate in the choice of one of a suitable value for this parameter, the function \textbf{sensitivity\_min\_SSU} allows performing a sensitivity analysis, showing how the first and second stage sample sizes vary by varying its values:

\begin{verbatim}
	> sens_min_SSU <- sensitivity_min_SSU (
	+             samp_frame=pop,
	+             id_PSU="municipality",
	+             id_SSU="id_ind",
	+             strata_var="stratum",
	+             target_vars=c("income_hh","active","inactive","unemployed"),
	+             deff_var="stratum",
	+             domain_var="region",
	+             minimum=50,
	+             delta=1,
	+             deff_sugg=1.5,
	+             min=30,
	+             max=80) 
\end{verbatim}

\begin{figure} [h!]
	\centering
	\includegraphics[width=14cm,height=10cm]{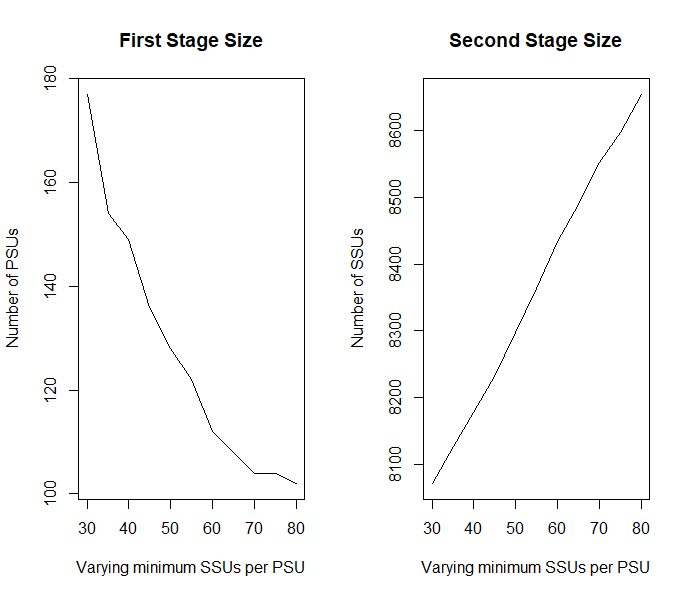}
	\caption{Sensitivity analysis for \textbf{minimum} parameter.}
	\label{PlotMinimum}
\end{figure}

This function calculates 10 different couples of values for the number of PSUs and SSU as resulting from the allocation step, starting with the value '30' assigned to the parameter \textbf{minimum}, ending with the value '80'.
The results are reported in Figure \ref{PlotMinimum}.

On the basis of the results of the sensitivity analysis, we can instantiate the required parameters, for example in this way:

\begin{verbatim}
	minimum = 50     # minimum number of SSUs to be interviewed in each selected PSU
\end{verbatim}
and execute the \textbf{prepareInputToAllocation1} function:

\begin{verbatim}
	> inp <- prepareInputToAllocation1(
	+             samp_frame = pop,
	+             id_PSU = "municipality",
	+             id_SSU = "id_ind",
	+             strata_var = "stratum",
	+             target_vars = c("income_hh","active","inactive","unemployed"),
	+             deff_var = "stratum",
	+             domain_var = "region",
	+             delta,     
	+             minimum)
	Computations are being done on population data
	Number of strata:  24
	... of which with only one unit:  0
\end{verbatim}

The output of this function (\textbf{inp}) is a list composed by the following elements:

\begin{enumerate}
	\item the \textbf{strata} dataframe
	\item the \textbf{deff} dataframe
	\item the \textbf{effst} dataframe
	\item the \textbf{rho} dataframe
	\item the \textbf{psu\_file} dataframe
	\item the \textbf{des\_file} dataframe
\end{enumerate}
that will be the inputs for the optimal allocation step (with the exception of the \textbf{deff}), which is produced only for documentation).

Here we report the content of the \textbf{rho} dataframe:
\begin{verbatim}
	  STRATUM RHO_AR1     RHO_NAR1 RHO_AR2         RHO_NAR2 RHO_AR3        RHO_NAR3
	1    1000       1 0.0032494875       1 0.00001260175649       1 0.0000003631192
	2    2000       1 0.0028554017       1 0.00150936389450       1 0.0007420929883
	3    3000       1 0.0069678726       1 0.00162968276279       1 0.0006469515878
	4    4000       1 0.0114552934       1 0.00578473329221       1 0.0019797687826
	5    5000       1 0.0002677333       1 0.00000001682475       1 0.0000029484212
	6    6000       1 0.0057050500       1 0.00004270905958       1 0.0000397945795
	  RHO_AR4       RHO_NAR4
	1       1 0.000039120880
	2       1 0.000937018761
	3       1 0.002837431259
	4       1 0.008962657055
	5       1 0.000003404961
	6       1 0.000194411580
\end{verbatim}
that has been calculated using the equation ~\eqref{eq:rho_pop}.

\subsubsection{Step 2: optimization of PSUs and SSUs allocation}

It is now possible to execute the optimization step of the sample design. 

First of all, we define the set of precision constraints on the target variables:
\begin{verbatim}
	   DOM  CV1  CV2  CV3  CV4
	1 DOM1 0.02 0.03 0.03 0.05
	2 DOM2 0.03 0.06 0.06 0.08
\end{verbatim}
We interpret the values of the CVs in this way: the maximum expected coefficient of variation for the first target variable (\textbf{household income}) is 2\% at the national level and 3\% at the regional level; for \textbf{active} and \textbf{inactive} the expected maximum values of CV is 3\% at the national level and 6\% at the regional level; finally, for \textbf{unemployed} it is 5\% at the national level and 8\% at the regional level.

The optimization step is performed by executing the \textbf{beat.2s} function:

\begin{verbatim}
> inp1$desfile$MINIMUM <- 50
> alloc1 <- beat.2st(stratif = inp1$strata, 
+                   errors = cv, 
+                   des_file = inp1$des_file, 
+                   psu_file = inp1$psu_file, 
+                   rho = inp1$rho, 
+                   deft_start = NULL,
+                   effst = inp1$effst, 
+                   minPSUstrat = 2,
+                   minnumstrat = 50
+                   )
  iterations PSU_SR PSU NSR PSU Total  SSU
1          0      0       0         0 7887
2          1     31     104       135 8328
3          2     39     104       143 8317
4          3     38     104       142 8320
\end{verbatim}

This design is characterized by 142 PSUs (of which 38 self-representative, SR, and 104 non self-representative, NSR) and 8,320 SSUs.

\subsubsection{Step 3: selection of PSUs and SSUs}

We can now proceed in selecting the PSUs:

\begin{verbatim}
> sample_1st <- select_PSU(alloc, type="ALLOC", pps=TRUE, plot=TRUE)
> sample_1st$PSU_stats
   STRATUM PSU PSU_SR PSU_NSR  SSU SSU_SR SSU_NSR
1     1000   2      2       0  286    286       0
2     2000   9      3       6  452    152     300
3     3000   4      0       4  200      0     200
4     4000   2      0       2  100      0     100
5     5000   2      2       0  219    219       0
6     6000   2      0       2  100      0     100
7     7000   2      0       2  100      0     100
8     8000   2      0       2  100      0     100
9     9000   1      1       0  557    557       0
10   10000   6      6       0  587    587       0
11   11000  26      2      24 1300    100    1200
12   12000   8      0       8  400      0     400
13   13000   1      1       0  703    703       0
14   14000   4      4       0  577    577       0
15   15000  27      9      18 1361    461     900
16   16000  18      0      18  900      0     900
17   17000   1      1       0  154    154       0
18   18000   4      2       2  200    100     100
19   19000   7      1       6  350     50     300
20   20000   4      0       4  200      0     200
21   21000   1      1       0  125    125       0
22   22000   3      3       0  150    150       0
23   23000   4      0       4  200      0     200
24   24000   2      0       2  100      0     100
25   Total 142     38     104 9421   4221    5200
\end{verbatim}

A discrepancy can be noted between the number of SSUs determined by the allocation step and the one produced by the selection of PSUs. This is because the selection of PSUs controls that the minimum number of SSUs to be allocated in each selected PSU is compliant with the minimum, in our case equal to 50: if not, this minimum is assigned. This is why the total number of SSUs increases from 8,320 to 9,421.

Selected PSUs are contained in the \textbf{sample\_PSU} element of the output list:
\begin{verbatim}
> head(sample_1st$sample_PSU)
  PSU_ID STRATUM stratum SR nSR PSU_final_sample_unit Pik weight_1st weight_2st   weight
1    330    1000  1000-1  1   0                   207   1          1   706.0966 706.0966
2    309    1000  1000-2  1   0                    72   1          1   706.1806 706.1806
3     51   10000 10000-0  1   0                   171   1          1   196.8480 196.8480
4     11   10000 10000-1  1   0                    96   1          1   196.9688 196.9688
5     40   10000 10000-2  1   0                    79   1          1   197.9494 197.9494
6     13   10000 10000-3  1   0                    72   1          1   198.3750 198.3750
\end{verbatim}

With this input, we can proceed to select the sample of final units:

\begin{verbatim}
> PSU_sampled <- sample_1st$sample_PSU
> samp <- select_SSU(df=pop,
+                    PSU_code="municipality",
+                    SSU_code="id_ind",
+                    PSU_sampled,
+                    verbose=TRUE)

PSU =  1  *** Selected SSU =  50
PSU =  4  *** Selected SSU =  72
PSU =  6  *** Selected SSU =  50
PSU =  8  *** Selected SSU =  557
...
PSU =  510  *** Selected SSU =  50
PSU =  512  *** Selected SSU =  50
--------------------------------
Total PSU =  142
Total SSU =  9421
--------------------------------
\end{verbatim}

\begin{figure} [h!]
	\centering
	\includegraphics[width=15cm,height=12cm]{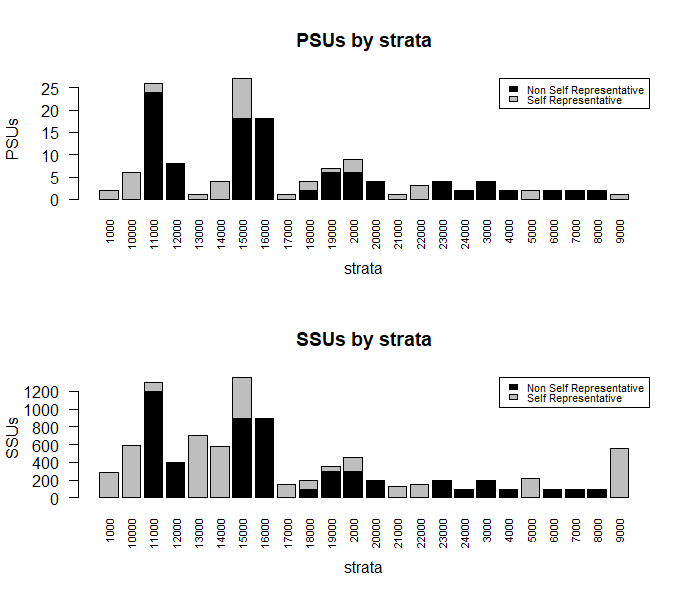}
	\label{fig:allocation1}
	\caption{Allocation of PSUs and SSUs (scenario 1).}
\end{figure}

The distribution of PSUs and SSUs in the different strata is reported in Figure~\ref{fig:allocation1}.

\begin{figure} [h!]
	\begin{subfigure}
		\centering
		\includegraphics[width=15cm,height=6cm]{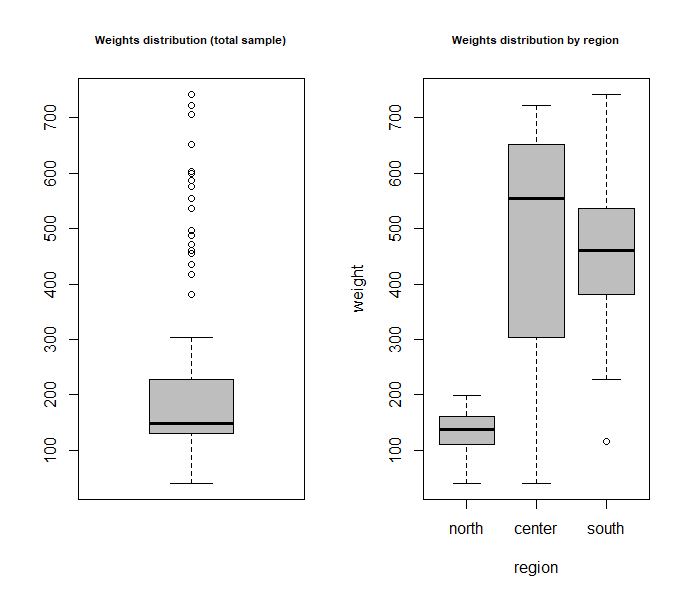}
		\label{weights11}
	\end{subfigure}
	\begin{subfigure}
		\centering
		\includegraphics[width=15cm,height=6cm]{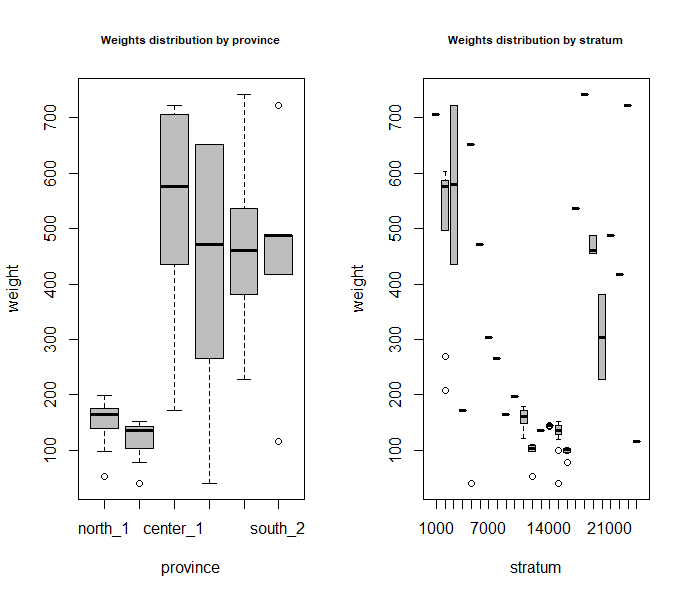}
		\label{weights12}
	\end{subfigure}
	\caption{Distribution of weights (scenario 1).}
	\label{weights1}
\end{figure}

In Figure \ref{weights1} the distribution of weights is reported.

It can be seen that the distribution of weights is variable at the national, regional and provincial level, and only inside each stratum, the variability is low, as desired, except for those strata in which for some PSUs the minimum number of SSUs (50) had to be attributed instead of the optimal allocation.

\subsubsection{Step 4: verify compliance with precision constraints}

The function \textbf{eval\_2stage]} allows verifying the compliance of the two-stage sample design to the set of precision constraints, by selecting a given number of different samples (in our case, 500) from the sampling frame, producing the estimates for each sample, and calculating over them the coefficients of variation for each target estimate.

We apply twice the function, first for the national level:

\begin{verbatim}
> # Domain level = national
> domain_var <- "one"
> set.seed(1234)
> eval11 <- eval_2stage(df,
+                     PSU_code,
+                     SSU_code,
+                     domain_var,
+                     target_vars,
+                     sample_1st$sample_PSU,
+                     nsampl=500, 
+                     writeFiles=FALSE,
+                     progress=FALSE) 
> eval11$coeff_var
     CV1    CV2    CV3    CV4  dom
1 0.0101 0.0091 0.0241 0.0344 DOM1
\end{verbatim}

then, at the regional level:

\begin{verbatim}
> # Domain level = regional
> domain_var <- "region"
> set.seed(1234)
> set.seed(1234)
> eval12 <- eval_2stage(df,
+                     PSU_code,
+                     SSU_code,
+                     domain_var,
+                     target_vars,
+                     sample_1st$sample_PSU,
+                     nsampl=500, 
+                     writeFiles=FALSE,
+                     progress=FALSE) 
> eval12$coeff_var
     CV1    CV2    CV3    CV4  dom
1 0.0113 0.0066 0.0235 0.0754 DOM1
2 0.0224 0.0206 0.0495 0.0733 DOM2
3 0.0240 0.0282 0.0515 0.0413 DOM3
\end{verbatim}

We recall that the precision constraints had been set equal to 2\% for the first variable, 3\% for the second and third, and 5\% for the fourth, at national level; and respectively to 3\% and 6\% and 8\% at regional level. We can see that the computed CVs are all compliant.

\subsection{Scenario 2 workflow}

Together with the availability of a sampling frame, containing the same information presented in the previous scenario, we assume also the availability of at least one previous round of the survey.
For sake of simplicity, we assume that the previous round sample is the same selected in scenario 1. We assume also that the values of the four target variables are the observed ones after the data collection. 
Having set the above conditions, the main difference with scenario 1 is that, instead of choosing in a somewhat arbitrarily way the values of the inputs required by the optimal allocation step, we can derive them directly from the collected survey data.

\subsubsection{Step 1: processing and analysis of survey data}

In this step, we proceed to perform the usual phases of calibration and production of the estimates. In doing that, we make use of the R package \textbf{ReGenesees}.

First we describe the sample design:

\begin{verbatim}
> ## Sample design description
> sample$stratum_2 <- as.factor(sample$stratum_2)
> sample.des <- e.svydesign(sample, 
+                           ids= ~ municipality + id_hh, 
+                           strata = ~ stratum_2, 
+                           weights = ~ weight,
+                           self.rep.str = ~ SR,
+                           check.data = TRUE)
\end{verbatim}

obtaining the \textbf{sample.des} object. Then we proceed with the calibration step:

\begin{verbatim}
> ## Calibration with known totals
> totals <- pop.template(sample.des,
+              calmodel = ~ sex : cl_age, 
+              partition = ~ region)
> totals <- fill.template(pop, totals, mem.frac = 10)
> sample.cal <- e.calibrate(sample.des, 
+                           totals,
+                           calmodel = ~ sex : cl_age, 
+                           partition = ~ region,
+                           calfun = "logit",
+                           bounds = c(0.3, 2.6), 
+                           aggregate.stage = 2,
+                           force = FALSE)
\end{verbatim}

obtaining the \textbf{sample.cal} object.

These two objects are what is needed to obtain, in an automated way, all the inputs required by the optimization step.

\subsubsection{Step 2: preparation of the inputs for the optimal sample design}

The preparation of all the inputs required by the optimization step is a straightforward operation by using the \textbf{prepareInputToAllocation2} function:

\begin{verbatim}
> inp <- prepareInputToAllocation2(
+         samp_frame = pop,             # sampling frame
+         RGdes = sample.des,           # ReGenesees design object
+         RGcal = sample.cal,           # ReGenesees calibrated object
+         id_PSU = "municipality",      # identification variable of PSUs
+         id_SSU = "id_hh",             # identification variable of SSUs
+         strata_vars = "stratum",      # strata variables
+         target_vars = c("income_hh","active","inactive","unemployed"), # target variables
+         deff_vars = "stratum",        # deff variables
+         domain_vars "region",         # domain variables
+         delta 0 1,                    # Average number of SSUs for each selection unit
+         minimum= 50                   # Minimum number of SSUs to be selected in each PSU
+       )
\end{verbatim}

The configuration of the output is just the same already seen in scenario 1 for the function \textbf{prepareInputToAllocation1}.

Here we report the content of the \textbf{effst} dataframe:

\begin{verbatim}
  stratum STRATUM   EFFST1    EFFST2    EFFST3    EFFST4
1    1000    1000 1.061891 0.9511291 0.9071854 1.0137193
2   10000   10000 1.005724 0.9077114 0.8991158 0.9780552
3   11000   11000 1.005722 0.9309392 0.9240808 0.9998968
4   12000   12000 1.026967 0.9241132 0.9117161 0.9911560
5   13000   13000 1.006354 0.9244961 0.9085689 0.9977077
6   14000   14000 1.002360 0.9348739 0.9237139 1.0065308
	...
\end{verbatim}

and of the \textbf{rho} dataframe:

\begin{verbatim}
  STRATUM RHO_AR1       RHO_NAR1 RHO_AR2        RHO_NAR2 RHO_AR3
1    1000       1 -0.00005314789       1  0.000004056338       1
2   10000       1  0.00021289157       1  0.000154688468       1
3   11000       1  0.01349102041       1 -0.004226612245       1
4   12000       1  0.00409179592       1  0.034025755102       1
5   13000       1  0.00002020513       1  0.000016396011       1
6   14000       1  0.00009018499       1 -0.000022736475       1
         RHO_NAR3 RHO_AR4      RHO_NAR4
1  0.000007542254       1 0.00003425352
2  0.000160791738       1 0.00002596213
3 -0.007398367347       1 0.00075012245
4  0.031294265306       1 0.02008032653
5  0.000019209402       1 0.00001038462
6 -0.000022157068       1 0.00007399651
	...
\end{verbatim}

in order to compare them with the scenario 1 ones.

\subsubsection{Step 3: optimization of PSUs and SSUs allocation}

The optimal allocation of PSUs and SSUs is the same as the one already seen in the first scenario: 

\begin{verbatim}
> set.seed(1234)
> inp2$des_file$MINIMUM <- 50
> alloc2 <- beat.2st(stratif = inp2$strata, 
+                   errors = cv, 
+                   des_file = inp2$des_file, 
+                   psu_file = inp2$psu_file, 
+                   rho = inp2$rho, 
+                   deft_start = NULL, 
+                   effst = inp2$effst,
+                   minnumstrat = 2, 
+                   minPSUstrat = 2)
  iterations PSU_SR PSU NSR PSU Total  SSU
1          0      0       0         0 9557
2          1     71      92       163 8464
3          2     38     108       146 8398
4          3     38     108       146 8396
\end{verbatim}

\subsubsection{Step 4: selection of PSUs and SSUs}

The selection of first and second stage units proceeds in exactly the same way than in the scenario 1, first selecting the PSUs, and then the SSUs. 

\begin{verbatim}
> sample_1st <- select_PSU(alloc2, type="ALLOC", pps=TRUE)
> sample_1st$PSU_stats
   STRATUM PSU PSU_SR PSU_NSR  SSU SSU_SR SSU_NSR
1     1000   2      2       0  279    279       0
2     2000  10      6       4  517    317     200
3     3000   4      0       4  200      0     200
4     4000   2      0       2  100      0     100
5     5000   2      2       0  202    202       0
6     6000   2      0       2  100      0     100
7     7000   2      0       2  100      0     100
8     8000   2      0       2  100      0     100
9     9000   1      1       0  564    564       0
10   10000   6      6       0  537    537       0
11   11000  26      4      22 1300    200    1100
12   12000  12      0      12  600      0     600
13   13000   1      1       0  756    756       0
14   14000   4      4       0  583    583       0
15   15000  28     10      18 1414    514     900
16   16000  22      0      22 1100      0    1100
17   17000   1      1       0  114    114       0
18   18000   2      0       2  100      0     100
19   19000   6      0       6  300      0     300
20   20000   4      0       4  200      0     200
21   21000   1      1       0  113    113       0
22   22000   2      0       2  100      0     100
23   23000   2      0       2  100      0     100
24   24000   2      0       2  100      0     100
25   Total 146     38     108 9579   4179    5400
>
> samp <- select_SSU(df=pop,
+                    PSU_code="municipality",
+                    SSU_code="id_ind",
+                    PSU_sampled=sample_1st$sample_PSU,
+                    verbose=TRUE)

PSU =  4  *** Selected SSU =  66
PSU =  8  *** Selected SSU =  564
PSU =  10  *** Selected SSU =  50
PSU =  11  *** Selected SSU =  96
...
PSU =  510  *** Selected SSU =  50
PSU =  512  *** Selected SSU =  50
--------------------------------
Total PSU =  146
Total SSU =  9579
--------------------------------
\end{verbatim}

\begin{figure} [h!]
	\centering
	\includegraphics[width=15cm,height=12cm]{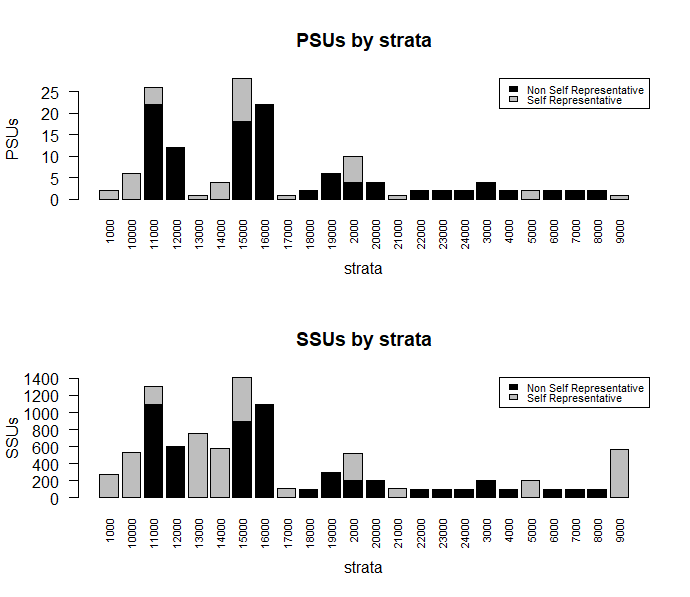}
	\label{allocation2}
	\caption{Allocation of PSUs and SSUs (scenario 2).}
\end{figure}

The distribution of PSUs and SSUs in the different strata is reported in Figure \ref{allocation2}. It can be seen that the relative distribution of both units in the strata is quite similar to the one already seen in scenario 1.

\begin{figure} [h!]
	\begin{subfigure}
		\centering
		\includegraphics[width=15cm,height=6cm]{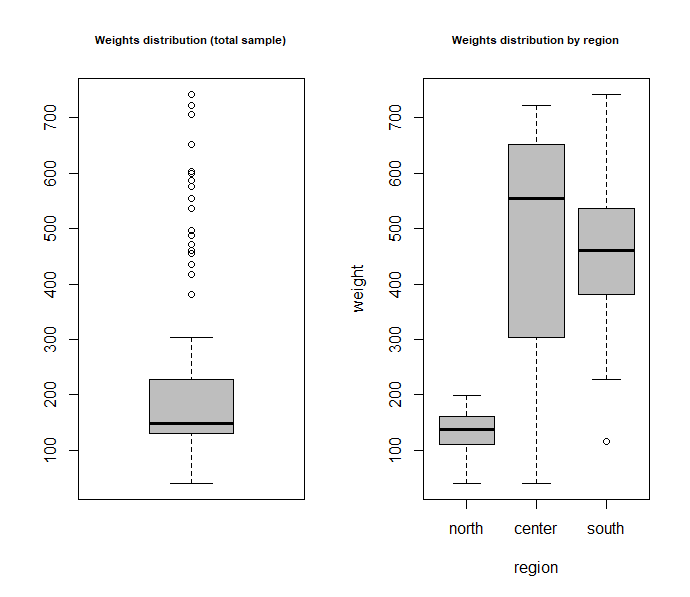}
		\label{weights21}
	\end{subfigure}
	\begin{subfigure}
		\centering
		\includegraphics[width=15cm,height=6cm]{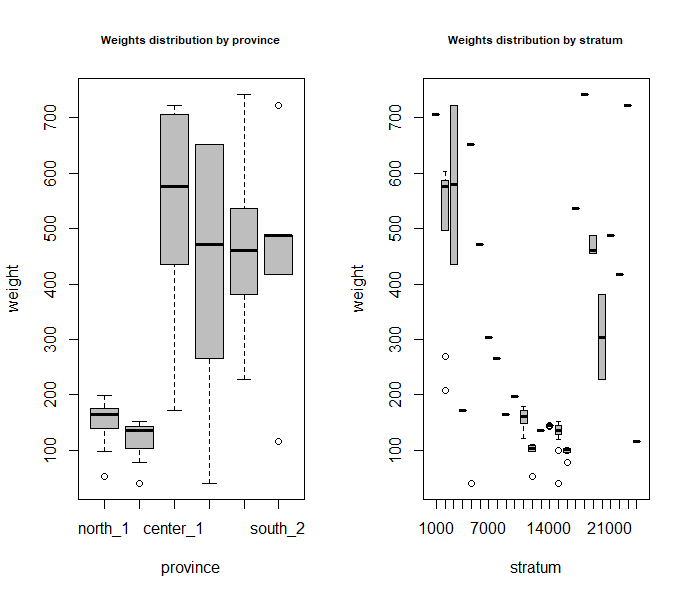}
		\label{weights22}
	\end{subfigure}
	\caption{Distribution of weights (scenario 2).}
	\label{weights2}
\end{figure}

We can observe now the distribution of weights in the selected sample (see Figure \ref{weights2}). Also in this case, their variability is lower inside the strata level, where it is almost null except in those strata in which for some PSUs the minimum number of SSUs (50) had to be attributed, instead of the optimal allocation.

\subsubsection{Step 5: verify the compliance to precision constraints}

As in the previous scenario, the final check consists in verifying the compliance of the optimized design to the precision constraints.

We, therefore, apply the function \textbf{eval\_2stage}, first for the national level:

\begin{verbatim}
> # Domain level = national
> domain_var <- "one"
> eval <- eval_2stage(df,
+                     PSU_code,
+                     SSU_code,
+                     domain_var,
+                     target_vars,
+                     PSU_sampled=sample_1st$sample_PSU,
+                     nsampl=500) 
> eval$coeff_var
    CV1    CV2   CV3    CV4  dom
1 0.012 0.0094 0.025 0.0364 DOM1
\end{verbatim}
then, at regional level:

\begin{verbatim}
> # Domain level = regional
> domain_var <- "region"
> eval <- eval_2stage(df,
+                     PSU_code,
+                     SSU_code,
+                     domain_var,
+                     target_vars,
+                     PSU_sampled=sample_1st$sample_PSU,
+                     nsampl=500) 
> eval$coeff_var
     CV1    CV2    CV3    CV4  dom
1 0.0105 0.0070 0.0246 0.0745 DOM1
2 0.0285 0.0206 0.0504 0.0748 DOM2
3 0.0291 0.0335 0.0597 0.0444 DOM3
\end{verbatim}

Also in this case, no precision constraint is violated. 

\section{Comparison with other softwares}

To evaluate the performance of R2BEAT, in this section we compare it to other two R packages, i.e.:

\begin{enumerate}
	\item the package \textbf{PracTools} \citep{practools} implements many of the procedures described in \cite{Valliant:2015}, including those regarding the design of multistage samples;
	\item the package \textbf{samplesize4surveys} \citep{Rojas:2020} allows to calculate the sample size for complex surveys.
\end{enumerate}

First, we briefly illustrate, for both packages, the functions covering the two-stage sampling design, then we apply them to the same case seen in scenario 1, finally comparing the obtained results
\footnote{In order to reproduce the processing related to the evaluation of the different softwares, datasets and R scripts are downloadable from the link  \href{https://github.com/barcaroli/Two-stage-sampling-software-comparison}{https://github.com/barcaroli/Two-stage-sampling-software-comparison}}.

\subsection{R package PracTools}

\cite{Valliant:2015} describe (pages 231-234) a method for the optimal allocation of two-stage sampling when numbers of sample PSUs and elements per PSU are adjustable (which is our case).

This method is implemented in the R function \textbf{clusOpt2} in the  \textbf{PracTools} package. This function computes the number of PSUs and the number of final units for each PSU for a two-stage sample which uses \textit{srs} at each stage or probability proportional to size with replacement (\textit{ppswr}) at the first stage and \textit{srs} at the second.

This function requires the indication of a number of parameters, among which:

\begin{itemize}
	\item C1: unit cost per PSU
	\item C2: unit cost per SSU
	\item delta: homogeneity measure 
	\item unit.rv: unit relvariance
	\item k: ratio of B2+W2 to unit relvariance
	\item CV0: target CV
	\item tot.cost: total budget for variable costs
	\item cal.sw: indicates if the optimization has to be run for a fixed total budget, or the for target CV0
\end{itemize}

The function \textbf{BW2stagePPS} computes the population values of B2, W2, and delta whose meaning is explained in \cite{Valliant:2015} (page 222).

The method is univariate: the optimization can be performed by indicating only one variable. The whole code required for the case described in scenario 1 is given here:

\begin{verbatim}
	> load("pop.RData")
	> library(PracTools)
	> # Probabilities of inclusion (I stage)
	> pp <- as.numeric(table(pop$municipality))/nrow(pop)
	> # variable income_hh
	> bw <- BW2stagePPS(pop$income_hh, pp, psuID=pop$municipality)
	> bw
	        B2          W2 unit relvar       B2+W2           k       delta 
	0.04075893  0.79538674  0.83601766  0.83614567  1.00015312  0.04874621 
	> des <- clusOpt2(C1=130,
	+                 C2=1,
	+                 delta=bw[6],
	+                 unit.rv=bw[3],
	+                 k=bw[5],
	+                 CV0=0.02,
	+                 tot.cost=NULL,
	+                 cal.sw=2)
	> des
	C1 = 130
	C2 = 1
	delta = 0.04874621
	unit relvar = 0.8360177
	k = 1.000153
	cost = 25499.72
	m.opt = 141.4
	n.opt = 50.4
	CV = 0.02
	> sample_size <- des$m.opt*des$n.opt
	> sample_size
	7126.56 
\end{verbatim}

In running the function, we have indicated that the optimization step was to be carried out having a target CV of 2\% for the variable \textbf{income\_hh}. As there is no way to directly indicate a desired minimum number of SSUs per PSU, we managed to obtain the desired value of 50 by indicating a couple of values 130 and 1 respectively for C1 and C2. As a result, the number of PSUs is 141 and the number of SSUs is 7,127.

\subsection{R package samplesize4surveys}

This package offers two functions to compute a grid of possible sample sizes for estimating single means (\textbf{ss2s4m}) or single proportions (\textbf{ss2s4p}) under two-stage sampling designs.

The required parameters are the following:

\begin{itemize}
	\item N: the population size
	\item mu: the value of the estimated mean of a variable of interest
	\item sigma: the value of the estimated standard deviation of a variable of interest
	\item conf: the statistical confidence
	\item delta: the maximum relative margin of error that can be allowed for the estimation
	\item M: number of clusters in the population
	\item to: (integer) maximum number of final units to be selected per cluster
	\item rho: the intraclass correlation coefficient
\end{itemize}

Here is the code we used in the case of the target variable \textbf{income\_hh}:

\begin{verbatim}
	> load("pop.RData")
	> PSU <- length(unique(pop$municipality))
	> pop_strata <- as.numeric(table(pop$stratum))
	> rho <- 0.04875369 # value taken from scenario 1 analysis
	> ss2s4m(N = nrow(pop), 
	+        mu = mean(pop$income_hh), 
	+        sigma = sd(pop$income_hh),
	+        delta = 0.02 * 1.96, 
	+        M = PSU, 
	+        to = 50, 
	+        rho = sum(rho$RHO_NAR1*pop_strata) / sum(pop_strata))
	50 3.388931 142 50 7061
\end{verbatim}

we obtain a design characterized by a total sample size of 7,061, with 142 PSUs.

Concerning the way we indicated the value of the parameter \textbf{rho}, we made use of the value of the intra-class correlation coefficient computed in scenario 1 by \textbf{R2BEAT}, not considering domains and strata. 

In order to compare the 2\% precision constrain expressed in terms of coefficient of variation, as the package requires the margin of error, we multiply the value of the CV by a z-value equal to 1.96, to obtain the ratio between the semi-width of the confidence interval and the estimate of the mean of the parameter. 

The use of the function \textbf{ss2s4p}, applicable for the other three variables, is practically the same.

\subsection{Comparison of results}

As already said, we refer to the scenario 1 setting. 

We consider the same precision levels for the four variables for the unique domain, set equals respectively to 2\%, 3\%, 3\% and 5\%.

We apply another constraint for all the three softwares, that is, we want to select a minimum number of final units in each PSU, set equal to 50. 

There is no problem in doing that for package \textbf{samplesize4surveys}, by setting the parameter \textbf{to} equal to 50: the last value of the final grid is the result we want. Moreover, there is no loss in the optimality of the solution in doing that, because the sample sizes obtained for further values are increasingly higher.

As for \textbf{PractTools}, it is more complicated because, as already said, there is no direct way to set this constraint. In any case, we manage to do that, by varying the value of C1 (leaving C2 equal to 1) until we find the solution with the nearest value of \textbf{n.opt} to 50.

A final consideration regarding the application of \textbf{R2BEAT}: in this setting, to be comparable with the other packages (that are univariate and mono-domain), it has been applied in a simplified way, that is, one variable per time (univariate), and no different domains and strata in the sampling frame. By so doing, \textbf{R2BEAT} yields obviously different results from those seen in scenario 1.

\begin{table}[h!]
	\caption{Two-stage sample design obtained by different packages.}
	\centering
	\begin{tabular}{|l||c|c|c|c|c|c| }
		\hline
		& \multicolumn{2}{c|}{PracTools} & \multicolumn{2}{c|}{R2BEAT} & \multicolumn{2}{c|}{samplesize4surveys} \\
		Variable   & PSUs &          SSUs           & PSUs &               SSUs               & PSUs &         SSUs         \\ \hline
		active     &  49  &          2459           &  37  &               2030               &  49  &         2436         \\
		inactive   &  90  &          4395           &  68  &               4338               &  88  &         4391         \\
		income\_hh & 141  &          7127           &  79  &               5140               & 142  &         7061         \\
		unemployed & 406  &         19956           & 149  &              10884               & 402  &        20058         \\ \hline
	\end{tabular}
	\label{results}
\end{table}	

\begin{figure} [h!]
	\begin{subfigure}
		\centering
		\includegraphics[width=12cm,height=7cm]{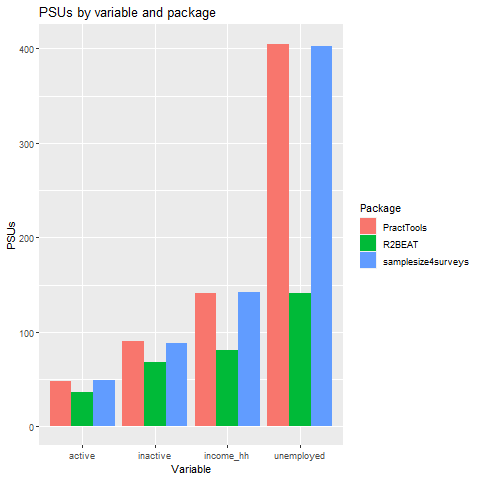}
		\label{comparison:fig1}
	\end{subfigure}
	\begin{subfigure}
		\centering
		\includegraphics[width=12cm,height=7cm]{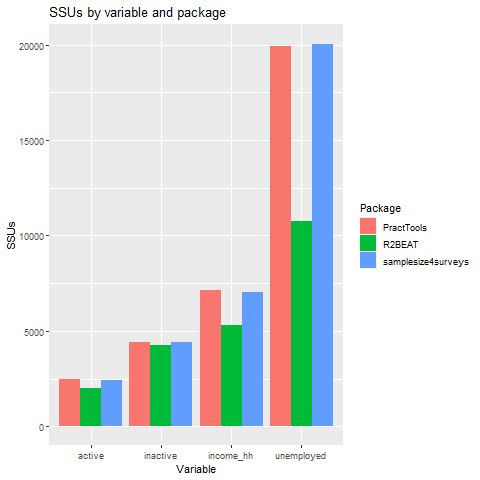}
		\label{comparison:fig2}
	\end{subfigure}
	\caption{Sample sizes by packages.}
	\label{comparison}
\end{figure}

In Table \ref{results} and in Figure \ref{comparison} are reported the results obtained by the three packages. To be sure of the results of \textbf{R2BEAT}, simulations have been carried out, and the resulting CVs are always below the precision threshold.

Analyzing the table, it is evident that \textbf{R2BEAT} is always the best performer, in terms of sample size, considering  both numbers of PSUs and SSUs.

\section{Concluding remarks}

Concluding, we would like to focus on the main strength of the R2BEAT, which could be considered its completeness, regarding all the phases of the statistical data production process.
The package deals with the design, stratification, allocation among strata and, finally, the selection of sample units. 

Furthermore, the facilities provided by R2BEAT are extremely flexible and generalizable: for instance, R2BEAT is the first R package on repositories to provide the optimal allocation, both for one-stage and two-stage sampling designs. 
These features make the package particularly helpful and valuable both for NSIs and private statistical institutions, such as marketing researchers, universities or national government organizations. 

Moreover, those who deal with statistics often have a large data availability, coming from registers, previous surveys or other data sources: the package, requiring auxiliary variables for designing and allocating the sample, makes this auxiliary information useful and profitable during the sampling planning process.

The output provided to users has been thought to be as clear as possible and to help them to carry out analysis and checks on the obtained allocations and on the sample on which the survey will be based.

Last, but not least, R2BEAT can be considered more efficient than the other available software and packages which deal with sample design: in fact, on equal errors, the sample size allocated is lower, both in terms of Primary and Secondary Stage Units (PSUs and SSUs.

\bibliographystyle{chicago}
\bibliography{refs}

\begin{thebibliography}{}

\bibitem[\protect\citeauthoryear{Baillargeon and Rivest}{Baillargeon and
  Rivest}{2011}]{Baillargeon+Rivest:2012}
Baillargeon, S. and L.-P. Rivest (2011).
\newblock The construction of stratified designs in r with the package
  stratification.
\newblock {\em Survey Methodology\/}~{\em 37\/}(1), 53--65.

\bibitem[\protect\citeauthoryear{Ballin and Barcaroli}{Ballin and
  Barcaroli}{2013}]{ballin2013joint}
Ballin, M. and G.~Barcaroli (2013).
\newblock Joint determination of optimal stratification and sample allocation
  using genetic algorithm.
\newblock {\em Survey Methodology\/}~{\em 39\/}(2), 369--393.

\bibitem[\protect\citeauthoryear{Barcaroli}{Barcaroli}{2014}]{barcaroli2014samplingstrata}
Barcaroli, G. (2014).
\newblock Samplingstrata: An r package for the optimization of stratified
  sampling.
\newblock {\em Journal of Statistical Software\/}~{\em 61\/}(4), 1--24.

\bibitem[\protect\citeauthoryear{Barcaroli, Buglielli, and Vitiis}{Barcaroli
  et~al.}{2020}]{maussr}
Barcaroli, G., T.~Buglielli, and C.~D. Vitiis (2020).
\newblock {\em {MAUSS-R}: {M}ultivariate {A}llocation of {U}nits in {S}ampling
  {S}urveys}.
\newblock R package version 2.4.

\bibitem[\protect\citeauthoryear{Bethel}{Bethel}{1985}]{bethel1985optimum}
Bethel, J.~W. (1985).
\newblock An optimum allocation algorithm for multivariate surveys.
\newblock In {\em Proceedings of the Social Statistics Section, ASA}, pp.\
  209--212.

\bibitem[\protect\citeauthoryear{Bethel}{Bethel}{1989}]{bethel1989sample}
Bethel, J.~W. (1989).
\newblock Sample allocation in multivariate surveys.
\newblock {\em Survey methodology\/}~{\em 15\/}(1), 47--57.

\bibitem[\protect\citeauthoryear{Biemer}{Biemer}{2010}]{biemer2010total}
Biemer, P.~P. (2010).
\newblock Total survey error: Design, implementation, and evaluation.
\newblock {\em Public Opinion Quarterly\/}~{\em 74\/}(5), 817--848.

\bibitem[\protect\citeauthoryear{Biemer and Lyberg}{Biemer and
  Lyberg}{2003}]{biemer2003introduction}
Biemer, P.~P. and L.~E. Lyberg (2003).
\newblock {\em Introduction to survey quality}, Volume 335.
\newblock John Wiley \& Sons.

\bibitem[\protect\citeauthoryear{Breidaks, Liberts, and Jukams}{Breidaks
  et~al.}{2020}]{surveyplanning}
Breidaks, J., M.~Liberts, and J.~Jukams (2020).
\newblock {\em {surveyplanning}: Survey planning tools}.
\newblock Riga, Latvia.
\newblock R package version 4.0.

\bibitem[\protect\citeauthoryear{Bueno}{Bueno}{2020}]{optimstrat}
Bueno, E. (2020).
\newblock {\em optimStrat: Choosing the Sample Strategy}.
\newblock R package version 2.3.

\bibitem[\protect\citeauthoryear{Chatterjee}{Chatterjee}{1968}]{chatterjee1968multivariate}
Chatterjee, S. (1968).
\newblock Multivariate stratified surveys.
\newblock {\em Journal of the American Statistical Association\/}~{\em
  63\/}(322), 530--534.

\bibitem[\protect\citeauthoryear{Chatterjee}{Chatterjee}{1972}]{chatterjee1972study}
Chatterjee, S. (1972).
\newblock A study of optimum allocation in multivariate stratified surveys.
\newblock {\em Scandinavian Actuarial Journal\/}~{\em 1972\/}(1), 73--80.

\bibitem[\protect\citeauthoryear{Choudhry, Rao, and Hidiroglou}{Choudhry
  et~al.}{2012}]{choudhry2012sample}
Choudhry, G.~H., J.~Rao, and M.~A. Hidiroglou (2012).
\newblock On sample allocation for efficient domain estimation.
\newblock {\em Survey methodology\/}~{\em 38\/}(1), 23--29.

\bibitem[\protect\citeauthoryear{Chromy}{Chromy}{1987}]{chromy1987design}
Chromy, J.~R. (1987).
\newblock Design optimization with multiple objectives.
\newblock {\em Proceedings of the Section on Survey Research Methods, 1987\/}.

\bibitem[\protect\citeauthoryear{Cicchitelli, Herzel, and
  Montanari}{Cicchitelli et~al.}{1992}]{cicchitelli1992campionamento}
Cicchitelli, G., A.~Herzel, and G.~E. Montanari (1992).
\newblock {\em Il campionamento statistico}.
\newblock Bologna: Il mulino.

\bibitem[\protect\citeauthoryear{Cochran}{Cochran}{1977}]{cochran1977sampling}
Cochran, W.~G. (1977).
\newblock {\em Sampling techniques}.
\newblock John Wiley \& Sons.

\bibitem[\protect\citeauthoryear{Conti and Marella}{Conti and
  Marella}{2012}]{conti2012campionamento}
Conti, P.~L. and D.~Marella (2012).
\newblock {\em Campionamento da popolazioni finite: Il disegno campionario}.
\newblock Springer Science \& Business Media.

\bibitem[\protect\citeauthoryear{Dalenius}{Dalenius}{1953}]{dalenius1953multi}
Dalenius, T. (1953).
\newblock The multi-variate sampling problem.
\newblock {\em Scandinavian Actuarial Journal\/}~{\em 1953\/}(sup1), 92--102.

\bibitem[\protect\citeauthoryear{Dalenius}{Dalenius}{1957}]{dalenius1957sampling}
Dalenius, T. (1957).
\newblock {\em Sampling in Sweden: contributions to the methods and theories of
  sample survey practice}.
\newblock Almqvist \& Wiksell.

\bibitem[\protect\citeauthoryear{Devaud and Till{\'e}}{Devaud and
  Till{\'e}}{2019}]{devaud2019deville}
Devaud, D. and Y.~Till{\'e} (2019).
\newblock Deville and s{\"a}rndal’s calibration: revisiting a 25-years-old
  successful optimization problem.
\newblock {\em TEST\/}~{\em 28\/}(4), 1033--1065.

\bibitem[\protect\citeauthoryear{Deville and S{\"a}rndal}{Deville and
  S{\"a}rndal}{1992}]{deville1992calibration}
Deville, J.-C. and C.-E. S{\"a}rndal (1992).
\newblock Calibration estimators in survey sampling.
\newblock {\em Journal of the American statistical Association\/}~{\em
  87\/}(418), 376--382.

\bibitem[\protect\citeauthoryear{Falorsi, Ballin, De~Vitiis, and Scepi}{Falorsi
  et~al.}{1998}]{falorsi1998principi}
Falorsi, P.~D., M.~Ballin, C.~De~Vitiis, and G.~Scepi (1998).
\newblock Principi e metodi del software generalizzato per la definizione del
  disegno di campionamento nelle indagini sulle imprese condotte dall’istat.
\newblock {\em Statistica Applicata\/}~{\em 10\/}(2), 235--257.

\bibitem[\protect\citeauthoryear{Folks and Antle}{Folks and
  Antle}{1965}]{folks1965optimum}
Folks, J.~L. and C.~E. Antle (1965).
\newblock Optimum allocation of sampling units to strata when there are r
  responses of interest.
\newblock {\em Journal of the American Statistical Association\/}~{\em
  60\/}(309), 225--233.

\bibitem[\protect\citeauthoryear{Gonzalez and Eltinge}{Gonzalez and
  Eltinge}{2010}]{gonzalez2010optimal}
Gonzalez, J.~M. and J.~L. Eltinge (2010).
\newblock Optimal survey design: A review.
\newblock {\em Section on Survey Research Methods - JSM\/}.
\newblock (Accessed on 15 March 2021).

\bibitem[\protect\citeauthoryear{Hartley}{Hartley}{1965}]{hartley1965multiple}
Hartley, H. (1965).
\newblock Multiple purpose optimum allocation in stratified sampling.
\newblock In {\em Proceedings of the Social Statistics Section, ASA}, pp.\
  258--261.

\bibitem[\protect\citeauthoryear{Horvitz and Thompson}{Horvitz and
  Thompson}{1952}]{horvitz1952generalization}
Horvitz, D.~G. and D.~J. Thompson (1952).
\newblock A generalization of sampling without replacement from a finite
  universe.
\newblock {\em Journal of the American statistical Association\/}~{\em
  47\/}(260), 663--685.

\bibitem[\protect\citeauthoryear{Huddleston, Claypool, and Hocking}{Huddleston
  et~al.}{1970}]{huddleston1970optimal}
Huddleston, H., P.~Claypool, and R.~Hocking (1970).
\newblock Optimal sample allocation to strata using convex programming.
\newblock {\em Journal of the Royal Statistical Society: Series C (Applied
  Statistics)\/}~{\em 19\/}(3), 273--278.

\bibitem[\protect\citeauthoryear{Kish}{Kish}{1965}]{kish1965survey}
Kish, L. (1965).
\newblock {\em Survey sampling}.
\newblock New York: John Wiley \& Sons, Inc.

\bibitem[\protect\citeauthoryear{Kish}{Kish}{1976}]{kish1976optima}
Kish, L. (1976).
\newblock Optima and proxima in linear sample designs.
\newblock {\em Journal of the Royal Statistical Society: Series A
  (General)\/}~{\em 139\/}(1), 80--95.

\bibitem[\protect\citeauthoryear{Kish}{Kish}{1988}]{kish1988multipurpose}
Kish, L. (1988).
\newblock Multipurpose sample designs.
\newblock {\em Survey Methodology\/}~{\em 14\/}(1), 19--32.

\bibitem[\protect\citeauthoryear{Kokan}{Kokan}{1963}]{kokan1963optimum}
Kokan, A. (1963).
\newblock Optimum allocation in multivariate surveys.
\newblock {\em Journal of the Royal Statistical Society: Series A
  (General)\/}~{\em 126\/}(4), 557--565.

\bibitem[\protect\citeauthoryear{Kokan and Khan}{Kokan and
  Khan}{1967}]{kokan1967optimum}
Kokan, A. and S.~Khan (1967).
\newblock Optimum allocation in multivariate surveys: An analytical solution.
\newblock {\em Journal of the Royal Statistical Society: Series B
  (Methodological)\/}~{\em 29\/}(1), 115--125.

\bibitem[\protect\citeauthoryear{Kozak, Verma, and Zielinski}{Kozak
  et~al.}{2007}]{kozak2007modern}
Kozak, M., M.~R. Verma, and A.~Zielinski (2007).
\newblock Modern approach to optimum stratification: Review and perspectives.
\newblock {\em Statistics in Transition\/}~{\em 8\/}(2), 223--250.

\bibitem[\protect\citeauthoryear{Kozak, Zieli{\'n}ski, and Singh}{Kozak
  et~al.}{2008}]{kozak2008stratified}
Kozak, M., A.~Zieli{\'n}ski, and S.~Singh (2008).
\newblock Stratified two-stage sampling in domains: Sample allocation between
  domains, strata, and sampling stages.
\newblock {\em Statistics \& probability letters\/}~{\em 78\/}(8), 970--974.

\bibitem[\protect\citeauthoryear{Neyman}{Neyman}{1934}]{neyman1934optimal}
Neyman, J. (1934).
\newblock On the two different aspects of the representative method: the method
  of stratified sampling and the method of purposive selection.
\newblock {\em Journal of the Royal Statistical Society\/}~{\em 97\/}(4),
  558--625.

\bibitem[\protect\citeauthoryear{Rojas}{Rojas}{2016}]{rojas2016estrategias}
Rojas, H. A.~G. (2016).
\newblock {\em Estrategias de muestreo: dise{\~n}o de encuestas y
  estimaci{\'o}n de par{\'a}metros}.
\newblock Ediciones de la U.

\bibitem[\protect\citeauthoryear{Rojas}{Rojas}{2020}]{Rojas:2020}
Rojas, H. A.~G. (2020).
\newblock {\em samplesize4surveys: Sample Size Calculations for Complex
  Surveys}.
\newblock R package version 4.1.1.

\bibitem[\protect\citeauthoryear{S{\"a}rndal}{S{\"a}rndal}{2007}]{sarndal2007calibration}
S{\"a}rndal, C.-E. (2007).
\newblock The calibration approach in survey theory and practice.
\newblock {\em Survey methodology\/}~{\em 33\/}(2), 99--119.

\bibitem[\protect\citeauthoryear{S{\"a}rndal, Swensson, and
  Wretman}{S{\"a}rndal et~al.}{2003}]{SSW:03}
S{\"a}rndal, C.-E., B.~Swensson, and J.~Wretman (2003).
\newblock {\em Model assisted survey sampling}.
\newblock Springer Science \& Business Media.

\bibitem[\protect\citeauthoryear{Stokes and Plummer}{Stokes and
  Plummer}{2004}]{stokes2004using}
Stokes, L. and J.~Plummer (2004).
\newblock Using spreadsheet solvers in sample design.
\newblock {\em Computational statistics \& data analysis\/}~{\em 44\/}(3),
  527--546.

\bibitem[\protect\citeauthoryear{Tillé and Matei}{Tillé and
  Matei}{2021}]{Rsampling}
Tillé, Y. and A.~Matei (2021).
\newblock {\em sampling: Survey Sampling}.
\newblock R package version 2.9.

\bibitem[\protect\citeauthoryear{Till{\'e}}{Till{\'e}}{2020}]{tille2020sampling}
Till{\'e}, Y. (2020).
\newblock {\em Sampling and estimation from finite populations}.
\newblock John Wiley \& Sons.

\bibitem[\protect\citeauthoryear{Tschprow}{Tschprow}{1923}]{tschprow1923optimal}
Tschprow, A. (1923).
\newblock On the two different aspects of the representative method: the method
  of stratified son the mathematical expectation of the moments of frequency
  distributions in the case of correlated observationsampling and the method of
  purposive selection.
\newblock {\em Metron\/}~{\em 2}, 646--683.

\bibitem[\protect\citeauthoryear{Valliant, Dever, and Kreute}{Valliant
  et~al.}{2015}]{Valliant:2015}
Valliant, R., J.~A. Dever, and F.~Kreute (2015).
\newblock {\em Practical Tools for Designing and Weighting Survey Samples}.
\newblock Springer.

\bibitem[\protect\citeauthoryear{Valliant, Dever, and Kreuter}{Valliant
  et~al.}{2020}]{practools}
Valliant, R., J.~A. Dever, and F.~Kreuter (2020).
\newblock {\em PracTools: Tools for Designing and Weighting Survey Samples}.
\newblock R package version 1.2.2.

\bibitem[\protect\citeauthoryear{Waters and Chester}{Waters and
  Chester}{1987}]{waters1987optimal}
Waters, J.~R. and A.~J. Chester (1987).
\newblock Optimal allocation in multivariate, two-stage sampling designs.
\newblock {\em The American Statistician\/}~{\em 41\/}(1), 46--50.

\bibitem[\protect\citeauthoryear{Yates}{Yates}{1960}]{yates1960sampling}
Yates, F. (1960).
\newblock Sampling methods for censuses and surveys , charles griffin and co.
\newblock {\em Ltd., London\/}.

\bibitem[\protect\citeauthoryear{Zardetto}{Zardetto}{2015}]{zardetto2015regenesees}
Zardetto, D. (2015).
\newblock Regenesees: An advanced r system for calibration, estimation and
  sampling error assessment in complex sample surveys.
\newblock {\em Journal of Official Statistics\/}~{\em 31\/}(2), 177--203.

\end{thebibliography}

\end{document}